\documentclass[twocolumn]{aastex61}
\usepackage{empheq}

\shorttitle{HST Imaging of Coma P}
\shortauthors{Brunker et al.}

\providecommand{\e}[1]{\ensuremath{\times 10^{#1}}}	
\accepted{December 28, 2018}
\submitjournal{AJ}

\begin{document}
\title{The Enigmatic (Almost) Dark Galaxy Coma P: Distance Measurement and Stellar Populations from HST Imaging\footnote{Based on observations made with the NASA/ESA \textit{Hubble Space Telescope}, obtained from the Data Archive at the Space Telescope Science Institute, which is operated by the Association of Universities for Research in Astronomy, Inc., under NASA contract NAS 5-26555.} }

\author[0000-0001-6776-2550]{Samantha W. Brunker}
\affiliation{Department of Astronomy, Indiana University, 727 East Third Street, Bloomington, IN 47405, USA}

\author[0000-0001-5538-2614]{Kristen B. W. McQuinn}
\affiliation{University of Texas at Austin, McDonald Observatory, 2515 Speedway, Stop C1402, Austin, TX 78712, USA}
 
 \author{John J. Salzer}
\affiliation{Department of Astronomy, Indiana University, 727 East Third Street, Bloomington, IN 47405, USA}

\author[0000-0002-1821-7019]{John M. Cannon}
\affiliation{Department of Physics \& Astronomy, Macalester College, 1600 Grand Avenue, Saint Paul, MN 55105, USA}

\author[0000-0001-9165-8905]{Steven Janowiecki}
\affiliation{Department of Astronomy, Indiana University, 727 East Third Street, Bloomington, IN 47405, USA}
\affiliation{International Centre for Radio Astronomy Research, University of Western Australia, 35 Stirling Highway, Crawley, WA 6009, Australia}

\author{Lukas Leisman}
\affiliation{Department of Physics and Astronomy, Valparaiso University, Valparaiso, IN 46383, USA}

\author{Katherine L. Rhode}
\affiliation{Department of Astronomy, Indiana University, 727 East Third Street, Bloomington, IN 47405, USA}

\author[0000-0002-9798-5111]{Elizabeth A. K. Adams}
\affiliation{ASTRON, the Netherlands Institute for Radio Astronomy, Postbus 2, 7990 AA, Dwingeloo, The Netherlands}
\affiliation{Kapteyn Astronomical Institute, University of Groningen, Postbus 800, 9700 AV, Groningen, The Netherlands}

\author[0000-0002-1895-0528]{Catherine Ball}
\affiliation{Department of Physics \& Astronomy, Macalester College, 1600 Grand Avenue, Saint Paul, MN 55105, USA}

\author{Andrew E. Dolphin}
\affiliation{Raytheon Company, PO Box 11337, Tucson, AZ 85734, USA}
\affiliation{Steward Observatory, University of Arizona, 933 North Cherry Avenue, Tucson, AZ 85721, USA}

\author{Riccardo Giovanelli}
\affiliation{Center for Astrophysics and Planetary Science, Space Sciences Building, 122 Sciences Drive, Cornell University, Ithaca, NY 14853, USA}

\author[0000-0001-5334-5166]{Martha P. Haynes}
\affiliation{Center for Astrophysics and Planetary Science, Space Sciences Building, 122 Sciences Drive, Cornell University, Ithaca, NY 14853, USA}

\correspondingauthor{Samantha W. Brunker}
\email{sbrunker@indiana.edu}

\begin{abstract}

We present \textit{Hubble Space Telescope (HST)} observations of the low surface brightness (SB) galaxy Coma P.  This system was first discovered in the Arecibo Legacy Fast ALFA \ion{H}{1} survey and was cataloged as an (almost) dark galaxy because it did not exhibit any obvious optical counterpart in the available survey data (e.g., Sloan Digital Sky Survey).  Subsequent WIYN pODI imaging revealed an ultra-low SB stellar component located at the center of the \ion{H}{1} detection. We use the \textit{HST} images to produce a deep color-magnitude diagram (CMD) of the resolved stellar population present in Coma P. We clearly detect a red stellar sequence that we interpret to be a red giant branch, and use it to infer a tip of the red giant branch (TRGB) distance of 5.50$^{+0.28}_{-0.53}$ Mpc. The new distance is substantially lower than earlier estimates and shows that Coma P is an extreme dwarf galaxy.  Our derived stellar mass is only 4.3 $\times$ 10$^5$ $M_\odot$, meaning that Coma P has an extreme \ion{H}{1}-to-stellar mass ratio of 81.  We present a detailed analysis of the galaxy environment within which Coma P resides. We hypothesize that Coma P formed within a local void and has spent most of its lifetime in a low-density environment.  Over time, the gravitational attraction of the galaxies located in the void wall has moved it to the edge, where it had a recent ``fly-by" interaction with M64. We investigate the possibility that Coma P is at a farther distance and conclude that the available data are best fit by a distance of 5.5 Mpc.
\end{abstract}

\section{Introduction}\label{sec:introduction}

It is notoriously difficult to detect and study low surface brightness (SB) galaxies.   As a consequence, they are typically underrepresented in optically selected samples of galaxies \citep{impey1997}.  This selection bias impacts studies of galaxy formation and evolution by leaving out an entire population of galaxies and/or an entire phase of galaxy evolution. These low SB galaxies often contain large amounts of atomic hydrogen, and the most extreme examples are likely to be identified by blind \ion{H}{1} surveys. 

The Arecibo Legacy Fast ALFA (ALFALFA) blind \ion{H}{1} survey \citep{giovanelli2005,haynes2011} has produced a sample of over 30,000 extragalactic \ion{H}{1} sources over 7000 square degrees of sky.  The vast majority of the extragalactic ALFALFA sources ($\sim$98\%) can be associated with optical counterparts in existing databases \citep[e.g., Sloan Digital Sky Survey (SDSS);][]{sdssDR14}, and most of the remaining objects are likely to be false detections or tidal debris.  A modest number of sources have no optical counterpart in available surveys, have velocities that place them at distances beyond the Local Group, and are sufficiently isolated that a tidal origin seems unlikely.  These sources have been named ``almost dark" galaxies, since most (but not all) have been found to possess extremely faint optical emission when imaged more deeply \citep[e.g,][]{cannon2015,steven2015,leisman2017}.  

The subject of this work, Coma P (AGC 229385), is a dwarf galaxy that is part of a system of ALFALFA neutral hydrogen detections designated HI1232+20.  The HI1232+20 system consists of three ALFALFA sources: AGC 229383, AGC 229384, and AGC 229385.  The first of these is resolved into two separate clouds in \ion{H}{1} interferometric maps \citep[see Figure 1 of][]{steven2015}.  The \ion{H}{1} sources in this system have significant amounts of gas ($>$10$\sigma$ ALFALFA detections) but are not detected in SDSS and lack obvious tidal companions.  All have radial velocities within 70 km/s of each other, are within 20$^\prime$ of each other on the sky, and appear to be isolated from other sources. An overlapping archival ultraviolet image from \textit{Galaxy Evolution Explorer} Data Release 7 \citep[GALEX DR7;][]{martin2005,morrissey2007,bianchi2014} shows a faint diffuse ultraviolet source at the location of Coma P.  

Deep \textit{gri} imaging of the HI1232+20 system with the pODI camera on the WIYN 3.5m at KPNO in 2013 revealed the ultra-low SB stellar component of Coma P \citep{steven2015}.  The galaxy exhibits a peak SB of $\mu_{g}$ = 26.4 mag/arcsec$^{2}$ and blue colors. A deep WIYN narrowband H$\alpha$ image revealed no detectable nebular emission.  The other ALFALFA sources in the HI1232+20 system showed no trace of any optical emission. 

The optical characteristics of Coma P presented in \cite{steven2015} are extreme, but when coupled with its \ion{H}{1} properties, this object becomes even stranger.  Based on that study, Coma~P falls far off all of the galaxy scaling relations of \citet{huang2012}. However, the interpretation of the properties of Coma P depend on knowing its distance with reasonable accuracy.  Coma P is located roughly 8\degr~north of the center of the Virgo Cluster, and the velocity flow model used for the ALFALFA catalog gives a preferred distance of 25 Mpc \citep{haynes2011}.  Flow-model distances are notoriously uncertain, however, particularly in the immediate vicinity of Virgo.  The ground-based optical and \ion{H}{1} data were sufficient to identify Coma P as extreme, but were insufficient to constrain its stellar properties and distance.  In order to determine the distance and to better understand the stellar populations of Coma P, our group obtained \textit{Hubble Space Telescope (HST)} observations which allow us to resolve the stellar component.  

In a companion study, \citet{Ball2018} studied the atomic interstellar medium of Coma P using a combination of Very Large Array (VLA) and Westerbork Synthesis Radio Telescope (WSRT) data.  \citet{Ball2018} found that Coma P is gas rich, but the \ion{H}{1} morphology is irregular.  The resolved stars in Coma P all appear to be coincident with the high column density \ion{H}{1} gas.  They conclude that the \ion{H}{1} morphology and kinematics are best described by two colliding \ion{H}{1} disks or a significant infall event.

In this work, we present results from our \textit{HST} optical imaging of Coma P, including resolved stellar photometry.  We measure the tip of the red giant branch (TRGB) distance to the galaxy, estimate the metallicity of the stellar populations, and constrain the recent star formation history (SFH). We also explore the environment within which Coma~P resides, and develop a scenario that explains most of the unusual features of this fascinating system.

\section{Observations and Photometry}\label{sec:obs}

\begin{figure*}[ht]
\epsscale{0.9}
\plotone{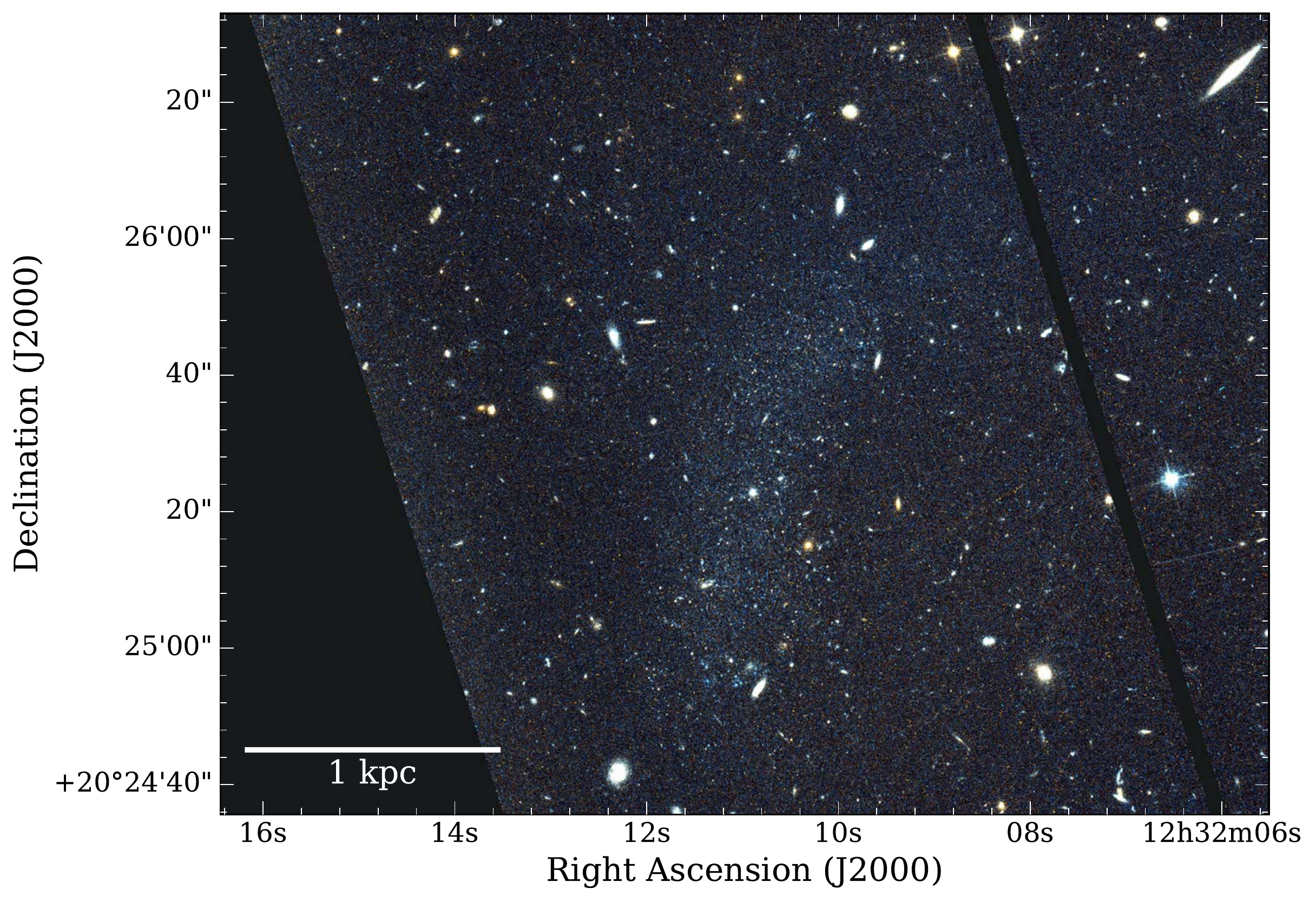}
\caption{\textit{HST} three-color image of Coma P combining the F606W image (blue), the average of F606W and F814W (green), and the F814W image (red).  North is up and east is left. The optical surface brightness of Coma P is approximately constant across the extent of the galaxy. Coma P appears to be elongated in the north$-$south direction, and does not have a regular morphology.} 
\label{fig:color}
\end{figure*}

The observations of Coma P were obtained with \textit{HST} using the Advanced Camera for Surveys (ACS) Wide Field Channel \citep{ford1998}. The observations were taken in two filters, F606W and F814W, with one \textit{HST} orbit per filter.  The exposures had total integration times of 2510 s in the F606W filter and 2648 s in the F814W filter.  The images were cosmic-ray cleaned and processed by the ACS pipeline.  The images were also corrected for charge transfer efficiency (CTE) nonlinearities. 

The pipeline-processed images were combined in AstroDrizzle \citep{Astrodrizzle} to produce the best quality images possible.  A color image of Coma P was made from the final drizzled images using the APLpy Python routine. Figure~\ref{fig:color} shows a zoomed-in region around Coma P.  This image confirms the extreme low SB nature of Coma P reported in \citet{steven2015}. As with the previous imaging study, the optical emission seems to have a roughly constant SB across the entire extent of the galaxy.  Coma P has an irregular morphology and is elongated in the north$-$south direction.   

Photometry was performed on the individual undrizzled flc.fits files with the ACS module of the HSTphot PSF-fitting photometry routine \citep{Dolphin2000}. DOLPHOT uses the transformations and uncertainties from \citet{Sirianni2005} to calibrate the photometry. Recent updates to the calibration constants result in values that are smaller by 1-3\% for the ACS broadband filters \citep{Bohlin2016}. This difference in calibration is well within our reported uncertainties for identifying the TRGB luminosities and does not have an impact on our conclusion. The photometry output list was filtered on several parameters measured for each detected source.  Only sources classified as stars with error flags $<$4 were kept. Cuts were also made on the sharpness and crowding parameters.  Sharpness indicates which sources are too broad (i.e., background galaxies) or too sharp (i.e., cosmic rays).  Crowding measures how much brighter a star would be if other stars nearby had not been fit simultaneously.  Higher values of crowding have higher photometric uncertainties. We rejected sources with ($V_{sharp}$ + $I_{sharp}$)$^{2}$ $>$ 0.075 and ($V_{crowd}$ + $I_{crowd}$) $>$ 0.8 \citep{mcquinn2014}. 

The photometry list was also filtered on the signal-to-noise ratios (S/Ns) in each filter.  To keep only the high-quality sources, we filtered both the F814W and F606W sources for S/N~$\ge$~5$\sigma$. Artificial star tests were performed, and the resulting sources were filtered on the same parameters in order to measure the completeness limit of the images.  The photometry lists were corrected for Galactic absorption ($A_{F606W}$ = 0.077 mag; $A_{F814W}$ = 0.048 mag) using the dust maps of \citet{Schlegel1998} with wavelength recalibration from \citet{Schlafly2011}. 

\begin{figure*}[ht]
\epsscale{0.59}
\plotone{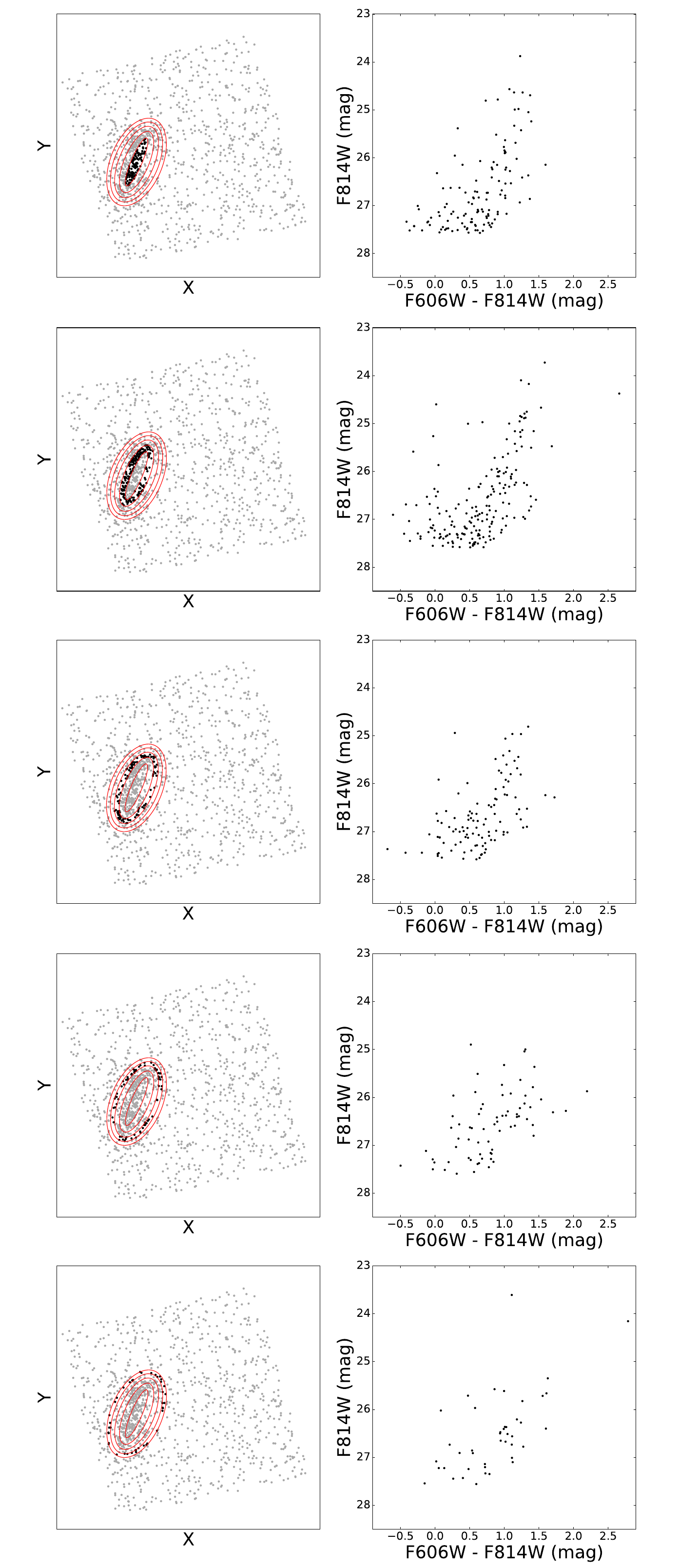}
\caption{CMDs plotted from concentric elliptical annuli centered on Coma P with ellipticities and position angles that approximate the stellar distribution of the galaxy. The left-hand panels show the spatial distribution of stars in gray, the ellipses in red, and the stars in each annulus are shown in black.  The right-hand panels show the CMDs corresponding to the specific annuli.  As the annuli move out farther from the center of the galaxy (top to bottom), they are sampling more background sources than stars in Coma P.  The outer two annuli (bottom two panels) are dominated by background sources and no longer show a distinct RGB.  We define the spatial extent of the galaxy as the third elliptical annulus shown (middle panel), which has a major axis of 78\arcsec, an ellipticity of 0.54, and a position angle of 157\degr.}
\label{fig:annuli_test}
\end{figure*}

\begin{figure*}[ht]
\epsscale{0.70}
\plotone{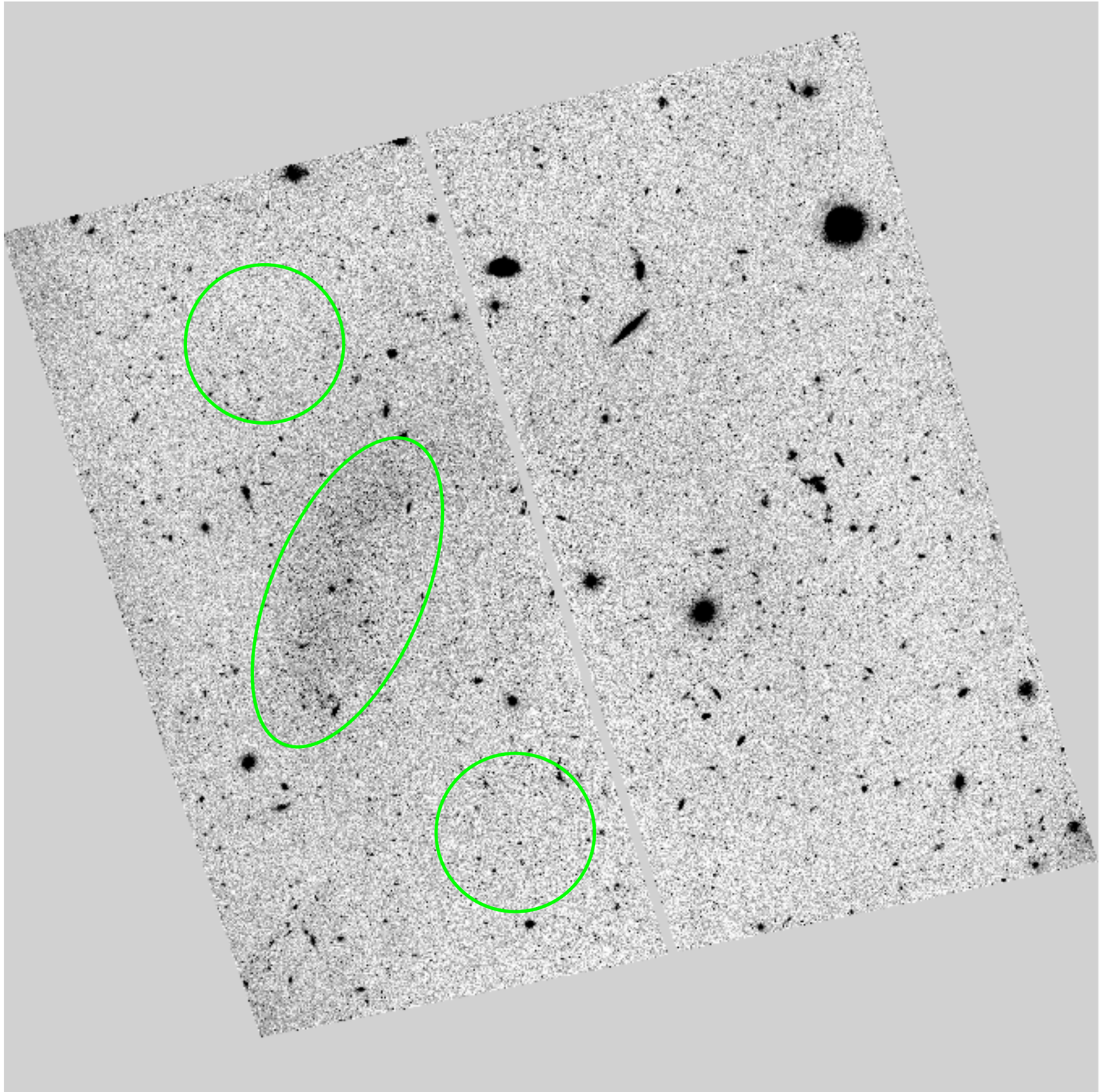}
\caption{Spatial extent of Coma P determined from Figure~\ref{fig:annuli_test} plotted on the drizzled F606W image. The elliptical aperture contains Coma P, but the galaxy has a more complex shape. Two circular regions whose combined areas are the same as the elliptical aperture are also plotted in order to compare the background to Coma P in Figure~\ref{fig:CMD_background}.}
\label{fig:final_region}
\end{figure*}

As seen in Figure~\ref{fig:color}, it is difficult to precisely identify the region occupied by Coma P because the galaxy is so faint and does not have a regular shape.  In order to determine the extent of the stellar component, we iteratively plotted the color-magnitude diagram (CMD) of point sources in concentric elliptic annuli centered on Coma P with ellipticities and position angles that approximately match the stellar distribution of the galaxy.  Ellipses could not be formally fit due to the low SB nature of Coma P, so we approximated the ellipticity and position angle of the galaxy by eye. Figure~\ref{fig:annuli_test} illustrates our procedure for establishing the size of the galaxy.  The ellipses are shown in red, and the stars in each annulus are shown in black.  As the annuli move out farther from the center of the galaxy (i.e., top to bottom in Figure~\ref{fig:annuli_test}), the red giant branch (RGB) shown in the corresponding CMD will become less distinct once the edge of the galaxy is reached.  The outer two annuli (bottom two panels on left side of Figure~\ref{fig:annuli_test}) show no clear RGB and are dominated mostly by background sources.  For this reason, we decided to define the outer edge of the galaxy as the third elliptical annulus shown in Figure ~\ref{fig:annuli_test} (middle panel), which has a major axis of 78\arcsec, an ellipticity of 0.54, and a position angle of 157\degr.  We adopt an uncertainty in the major axis of 5\arcsec, the average size of the increment between the ellipses shown in Figure ~\ref{fig:annuli_test}.  

\begin{figure*}[ht]
\epsscale{0.85}
\plotone{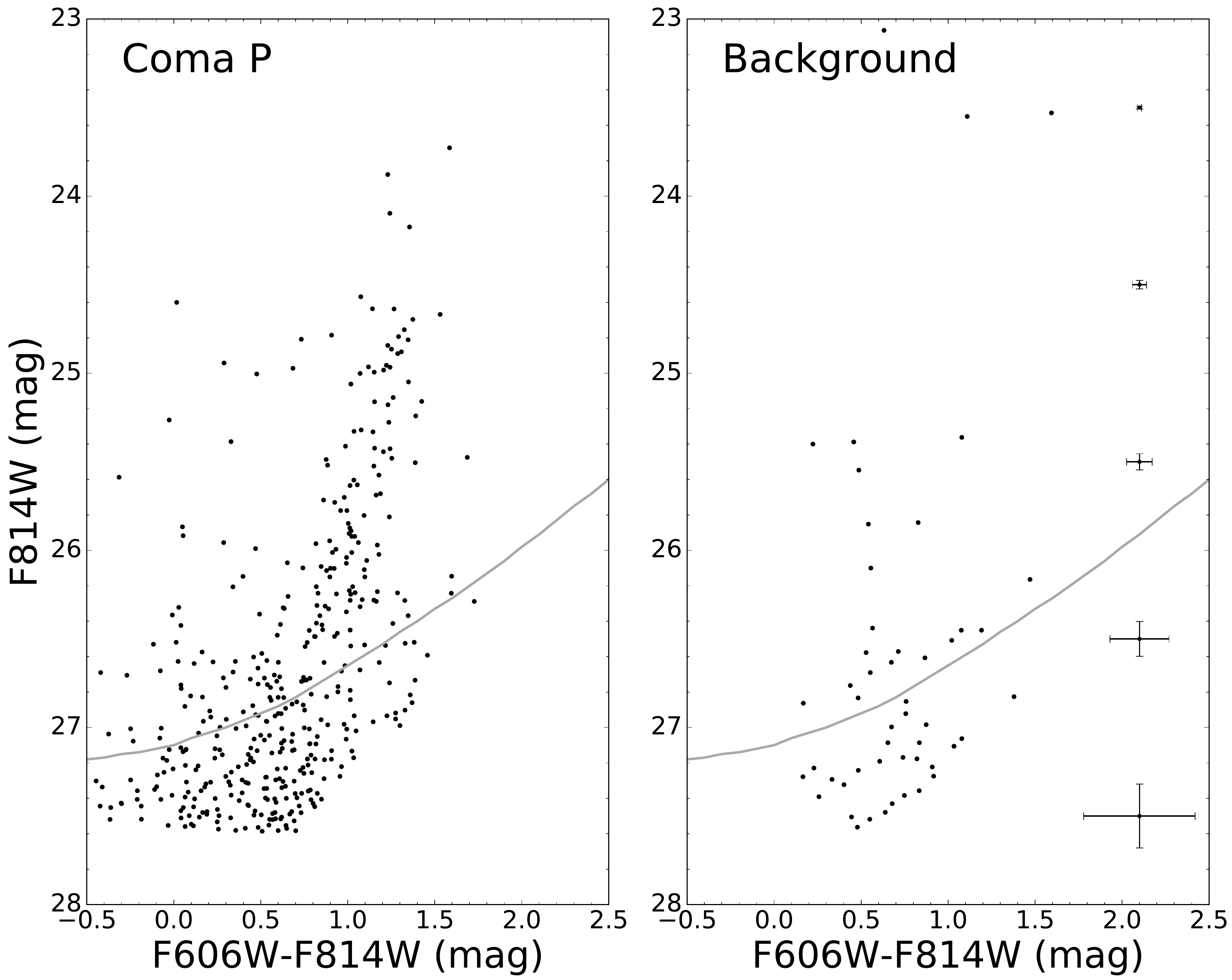}
\caption{Left: CMD of Coma P constructed from resolved stellar photometry from HST ACS imaging.  There are 420 stars in this CMD.  The photometry was corrected for Galactic absorption \citep{Schlafly2011}.  The 50\% completeness curve determined from artificial star tests is plotted as a solid gray line.  The RGB is identified in the CMD, but there is no indication of a well-populated upper main sequence.  Right: CMD of the background sources from circular regions identified in Figure~\ref{fig:final_region} which are located on the same CCD chip as Coma P but do not overlap the region occupied by the galaxy.  Representative uncertainties are plotted per magnitude bin in F814W.}
\label{fig:CMD_background}
\end{figure*}

The ellipse that defines the spatial extent of Coma P is plotted on the final F606W image in Figure~\ref{fig:final_region}. Two circular background regions that together cover the same area as the galaxy were selected and are also shown in Figure~\ref{fig:final_region}.  The overdensity of stars associated with the galaxy appears to be contained entirely within our final ellipse, but it is clear that Coma P is not precisely elliptical in shape.  

The left panel of Figure~\ref{fig:CMD_background} shows a CMD of the sources that lie within the elliptical region of Coma P shown in green on Figure~\ref{fig:final_region}. The 50\% completeness curve, based on our artificial star tests, is also shown.  The curve is derived by convolving the completeness functions for the F606W and F814W filters for a full range of colors. The data reach 50\% completeness at 27.2 mag for F814W and 28.4 for F606W.  The sources located within the two circular background regions are shown in the right panel of Figure~\ref{fig:CMD_background} along with representative photometric uncertainties as a function of F814W magnitude. There is a clear RGB in the Coma P CMD, but no strong presence of upper main-sequence stars is detected.  There are no clear features visible in the CMD of the background data.  

\section{TRGB Distance Determination}\label{sec:distance}

The presence of the RGB in the Coma P CMD allows us to use the TRGB method to measure the distance to the galaxy.  The TRGB distance method utilizes the \textit{I}-band luminosity of the brightest RGB stars, and it is fairly insensitive to metallicity and age.  The TRGB method is considered a standard candle distance measurement because the \textit{I}-band luminosity of low-mass stars prior to the helium flash is stable and predictable.  The TRGB distance method coupled with \textit{HST} imaging has produced accurate distances to many galaxies within and around the Local Volume \citep[e.g.,][]{sakai1997,kara2003,mcconn2004,Rizzi2007,dalcanton2009,tully2009,mcquinn2014}. 

A discontinuity in the \textit{I}-band luminosity function (LF) is what identifies the TRGB luminosity. Identifying the discontinuity requires observations of resolved stellar populations in the \textit{V} and \textit{I} bands that reach $\geqslant$1 mag below the TRGB in the \textit{I}-band. 
Both filters are needed in order to separate RGB stars in the CMD based on their color; the photometric depth is needed in order to identify the break in the LF that corresponds to the TRGB with high accuracy.  

For the TRGB measurement, the photometry has been transformed with a color term to account for metallicity effects.  This metallicity correction ``straightens'' the RGB, enabling a more precision measurement of the TRGB luminosity \citep[e.g.,][]{madore2009}. We use a TRGB luminosity calibration from \citet{Rizzi2007} for the ACS filters,
\begin{eqnarray}
M_{F814W}^{ACS} = &-&4.06(\pm 0.02) \nonumber \\ &+&~0.20(\pm 0.01) \cdot [(F606W - F814W) - 1.23]. \nonumber \\
\label{eq:trgb}
\end{eqnarray}

Note that a revised calibration for the TRGB luminosity in the \textit{HST} ACS filters that provides a higher order, quadratic correction for metallicity based on observations that cover a wide range of RGB colors has recently been made \citep[i.e., a \textit{V-I} color range from 1.4$-$4.3 mag;][]{Jang2017}. For the modest color range of the RGB stars in Coma~P, the revised calibration translates to a distance modulus within 0.02 mag of our reported values using the calibration from \citet{Rizzi2007}, which is within the uncertainties of the zero points of the two calibrations.

We used a Bayesian maximum likelihood (ML) method to identify the discontinuity in the LF. The ML technique is considered to be reliable and robust as it takes into account the photometric uncertainties and completeness of the data measured by the artificial star tests. This is particularly important to accurately identify the TRGB in cases where the data reach only $\sim1$ mag below the TRGB and/or in low-mass systems that have fewer stars populating the RGB.  Our photometry for Coma P reaches more than $\sim$2 mag below the TRGB (Figure~\ref{fig:CMD_background}), but given that the CMD has a relatively small number of stars ($\sim500$), we employ the ML method as our primary TRGB detection technique. We used a Sobel filter edge-detection algorithm as a check to the ML result.

\begin{figure*}[ht]
\epsscale{0.51}
\plotone{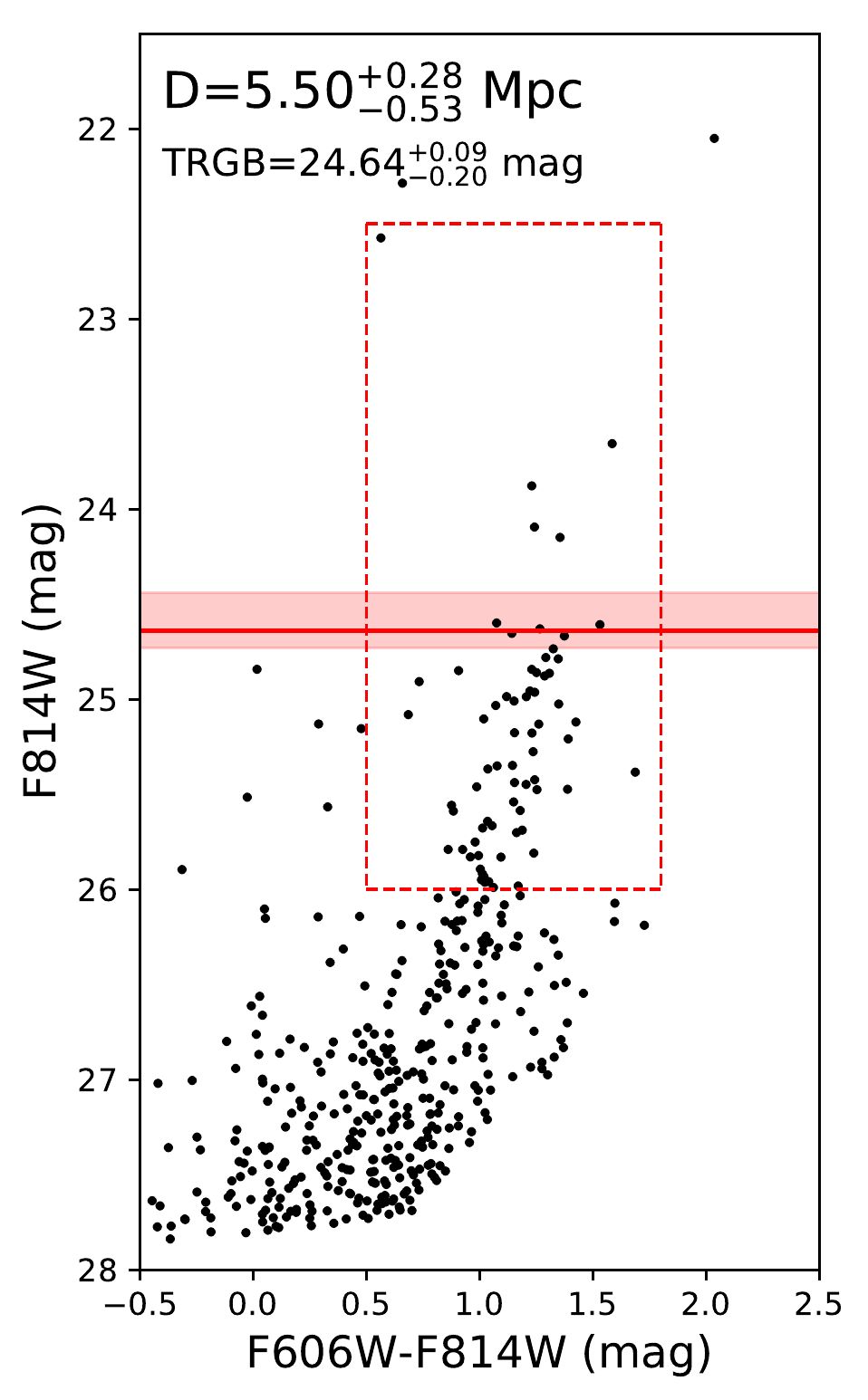}
\caption{CMD of Coma P using sources within the ellipse shown in Figure~\ref{fig:final_region}. For the TRGB measurement, the photometry has been transformed with a color term to account for metallicity effects. The RGB is clearly detected, and the dashed red line identifies the portion of RGB stars used in the TRGB measurement. The selection of stars is restricted to the brighter part of the RGB to avoid the larger uncertainties associated with the fainter stars, and it is restricted in color space to avoid including non-RGB stars.  The solid red line marks the luminosity of the TRGB identified by the maximum likelihood technique, and the shaded region indicates the uncertainties. The TRGB luminosity and the final distance measurement for Coma P are stated at the top.}
\label{fig:CMD_TRGB}
\end{figure*}

For both techniques, the RGB stars need to be selected from the extinction-corrected colors and magnitudes. Once those sources are identified, only the F814W LF is used  to measure the TRGB luminosity.  Figure~\ref{fig:CMD_TRGB} shows the CMD for Coma P used for our distance determination. The red dashed rectangle encompasses the stars used in the determination of the TRGB luminosity. Before applying either technique to measure the TRGB luminosity, the F814W LF of the selected RGB stars was transformed for metallicity effects using the color calibration from \citet[][Equation~(\ref{eq:trgb})]{Rizzi2007}. The final zero-point calibration was used with the measured TRGB luminosities to obtain the distance modulus from each technique.

\subsection{ML Technique}\label{sec:ml}
The ML method determines the TRGB luminosity by fitting a parametric form to the F814W LF transformed for metallicity effects \citep[e.g.,][]{sandage1979, mendez2002, makarov2006}. We assumed the following theoretical shape of the RGB LF from \citet{makarov2006}:

\begin{subequations}
\begin{empheq}[left={P = }\empheqlbrace]{alignat=2}
	& 10^{(A*(m - m_{TRGB}) + B)}, & \quad \text{if m - m$_{TRGB} \geqslant 0$}\\
	& 10^{(C*(m - m_{TRGB}))}, & \quad \text{if m - m$_{TRGB} < 0$}
\end{empheq}
\label{eq:ml_form}
\end{subequations}

\noindent where \textit{A} is the slope of the RGB with a normal prior of 0.30 and $\sigma=0.07$, \textit{C} is the slope of the asymptotic giant branch (AGB) with a normal prior of 0.30 and $\sigma=0.2$, \textit{B} is the RGB jump, and all three are treated as free parameters. The range in solutions returning log \textit{P} within 0.5 of the maximum gives the uncertainty, assuming a normal distribution. The TRGB luminosity from the ML technique is $24.64\pm0.09$ mag. 

\subsection{Sobel Filter}\label{sec:sobel}

The Sobel filter method was used as a check of the ML method.  The Sobel method determines the TRGB luminosity by identifying the break in the F814W LF using a Sobel edge-detection filter with a kernel of $[-2,~0,~2]$ \citep{lee1993,sakai1996,sakai1997}.  The bin width of the luminosity function was varied from 0.05 to 0.20 in steps of 0.05, and a bin width of 0.10 was determined to yield the best results. 

The Sobel responses often have multiple significant peaks.  In order to determine which peak corresponded to the TRGB discontinuity, we checked each response against the data in the CMD.  The TRGB luminosity from the Sobel filter technique is 24.60 mag, which is very close to the value determined by the ML method.

\subsection{Estimation of Uncertainties}\label{sec:results}

Overall, the results from the ML and Sobel filter methods agree quite well within the uncertainties. The Sobel filter response returned a brighter magnitude (smaller distance) for the TRGB, which is likely due to the binning of the LF or the potential presence of a non-RGB population.

Because the ML technique does not rely on binning and takes into account the photometric uncertainties and completeness, we adopt the TRGB luminosity from the ML technique. We apply the zero-point calibration from \citet{Rizzi2007} in Equation~\ref{eq:trgb} to calculate the final distance modulus of 28.70 mag. The statistical uncertainty is based on adding in quadrature the uncertainties from the TRGB zero-point calibration ($\sigma = 0.02$), the color-dependent metallicity correction ($\sigma = 0.01$), and the ML uncertainties calculated from the probability distribution function, which include uncertainties from the photometry and artificial star tests. Thus, the total statistical uncertainty is $\pm$0.09 mag. 

Given the low-mass nature of Coma P, the uncertainties on the TRGB luminosity (and, hence, the distance modulus) may be underestimated. If the RGB in a galaxy is sparsely populated, then the identified break in the LF may correspond to a grouping of the brightest RGB stars in the system rather than to the actual TRGB. For example, in the very low-mass galaxy Leo~P \citep{giovanelli2013}, based on the break in the \textit{I}-band LF at 22.11 mag, the distance modulus was calculated to be 26.19$^{+0.17}_{-0.5}$ mag from ground-based imaging of the resolved stellar populations; the lower uncertainties were based on modeling CMDs of galaxies with similarly low stellar mass and take into account the potentially underpopulated TRGB region \citep{mcquinn2013}. The distance modulus was later revised downward to be 26.05$\pm{0.20}$ mag using horizontal branch stars and RR Lyrae stars as distance indicators \citep{mcquinn2015a}, confirming that the identified break in the LF was slightly below the actual TRGB. Similar findings have been reported for the very low-mass galaxy Leo~A in the Local Group \citep{dolphin2002,Cole2007}. 

Thus, to conservatively estimate the lower uncertainties on the distance modulus, we consider the possibility that the upper RGB in Coma~P may be underpopulated. We compare the CMD in Figure~\ref{fig:CMD_TRGB} to previous CMDs and modeling of low-mass galaxies. For Leo~P, \citet{mcquinn2013} estimated the uncertainties on the TRGB luminosity to be $0.3-0.4$ mag for a CMD with $\sim40$ stars within 1 mag of the TRGB. Higher uncertainties of 0.5 mag were adopted for Leo~P based on the small grouping of stars in the CMD at brighter magnitudes. Similarly, \citet{makarov2006} found uncertainties to be of order 0.20 mag for a CMD with $\sim50$ stars within 1 mag of the TRGB. The CMD of Coma~P has 53 stars within 1 mag of the identified break in the LF, with no additional stars within $>0.5$ mag at brighter magnitudes. We adopt the higher 0.20 mag from \citet{makarov2006} as the lower uncertainty for the TRGB luminosity and distance modulus.

Systematic uncertainties are estimated to be $\sigma = 0.07$ mag, based on the calibration of the TRGB luminosity to the distance scale determined from the Large Magellanic Cloud \citep{carretta2000, Rizzi2007}. The final distance modulus from the ML TRGB luminosity with both statistical and systematic uncertainties is $28.70^{+0.09}_{-0.20}\pm0.07$ or, combined in quadrature, $28.70^{+0.11}_{-0.21}$ mag. This results in a distance to Coma P of 5.50$^{+0.28}_{-0.53}$ Mpc.

\begin{figure*}[ht]
\epsscale{1.1}
\plotone{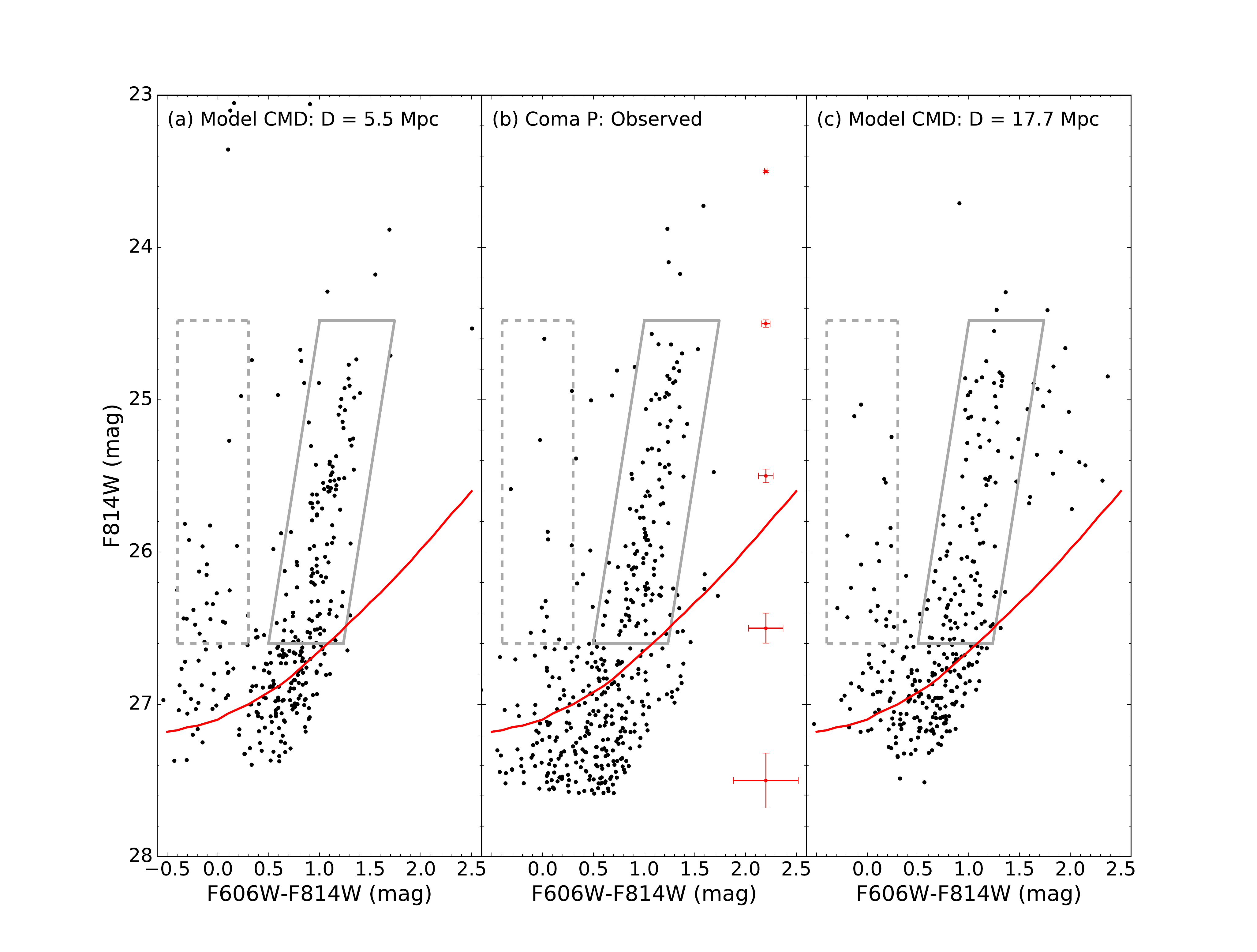}
\caption{CMD comparison between stellar population models for two possible distances to Coma P and the observed \textit{HST} data.  Panel (a) shows the model CMD for an assumed distance of 5.5 Mpc, the value adopted in this paper based on the TRGB-fitting process.  Panel (b) shows the observed CMD (same data as the left panel in Figure~\ref{fig:CMD_background}).  Panel (c) presents a model CMD assuming a distance of 17.7 Mpc, as described in the text.  Two regions that denote the blue-sequence region (dashed lines) and the red-sequence region (solid lines) are delimited.  The location of these regions is identical in all three panels.}
\label{fig:cmdmodel}
\end{figure*}

\subsection{Search for Additional Alternate Distances}

Due to the large difference in the distance to Coma P used by \citet[][25 Mpc]{steven2015} and the distance derived above, we explored the possibility that our TRGB distance might be incorrect.   In particular, we wanted to check whether the observed CMD could be explained by a larger distance, where the TRGB was located at such a faint magnitude that it was not clearly visible in our data.  We used stellar evolution isochrones and the CMD-fitting code {\sc match}, described below (\S \ref{sec:sfh}), to test whether farther distances could explain the observed composite stellar population in the CMD. This approach does not depend on fitting the discontinuity of the TRGB, but on how well the total number of point sources in a CMD are fit by the stellar models assuming different distances.  It is worth stressing that the accuracy of this method depends critically on how well the modeled SFHs match the actual system being analyzed.  This is particularly true for a low-mass galaxy like Coma~P. 

This process resulted in the best match between the observed and modeled CMDs for a distance of $\sim$17.7 Mpc.  At this distance, the TRGB would be located at F814W = 27.2, below the 50\% completeness limit of the current \textit{HST} data.  If Coma P were at this distance, the red-sequence stars seen in the CMD would be red supergiant stars rather than RGB stars.  A significant population of young stars would be required to explain the observed numbers of red supergiants.

A model CMD for this larger distance is shown in the rightmost panel of Figure \ref{fig:cmdmodel}, along with the observed CMD for Coma P and a model CMD derived using the same software but assuming a distance of 5.50 Mpc and the SFH presented in Section \ref{sec:sfh}.  In all three CMDs, we delimit two regions: a region we will refer to as the blue sequence (dashed lines) and a region we will refer to as the red sequence (solid lines).  We stress that the precise locations of the demarcation lines for these two regions are arbitrary; they are largely meant to guide the eye.  The 50\% completeness curve from Figure~\ref{fig:CMD_background} is reproduced in all three panels.  The overall similarities of the three CMDs are apparent.  In particular, the numbers of stars located along the red sequences is similar in number and distribution in all three plots.  The number of stars located in the blue-sequence region of the 5.5 Mpc model is the same as in the observed CMD, while the 17.7 Mpc model includes $\sim$70\% more blue-sequence stars.  We do not place strong significance in this difference, however, since the number of stars is small.

The most dramatic difference between the two models and the observed CMD is the presence of over a dozen luminous red stars located redward of the red-squence region in the 17.7 Mpc model.  The observed CMD has only one star in this same region (i.e., to the right of the red-sequence region and brighter than F814W = 25.8), while the 5.5 Mpc model based on the PARSEC library has an expectation value of two to three stars in this part of the CMD. Note that a different choice of stellar evolution isochrones will impact the number of red stars predicted.  The presence of these luminous red supergiants is a consequence of the high level of recent star formation (i.e., within the last several hundred Myr) required for the 17.7 Mpc model to have a strongly populated red sequence. We note that these luminous red stars are actually observed in low-mass galaxies with significant amounts of recent star formation.  The recent study of KDG 215 by \citet{Cannon2018} reveals a CMD with a substantial population of luminous stars redward of F606W~$-$~F814W = 1.5.  KDG 215 has a stellar mass comparable to what Coma P would have if the latter galaxy were located at a distance of 17.7 Mpc. 

Adopting the larger distance to Coma P is appealing, because it places it at a distance much closer to the value derived from the velocity flow model used by \citet{steven2015}.  In fact, this distance places Coma P in the outskirts of the Virgo Cluster. However, the nearly complete lack of luminous stars redward of the red sequence in the observed CMD does not agree with the models or observations of similar systems. Furthermore, the larger distance increases the deviation of Coma~P in galaxy scaling relations.  For example, Coma~P would lie a factor of $\sim$50 above the fiducial baryonic Tully$-$Fisher relation \citep[BTFR;][]{mcgaugh2012} if the 17.7 Mpc distance were adopted \citep[see Figure 17 of][]{Ball2018}.  The current HST data do not make it possible to rule out the larger distance with certainty. However, the balance of the evidence tends to support the 5.50 Mpc distance.

\subsection{Alternate Distance Proposed in \citet{anand2018}}\label{sec:tully}

After the initial submission of this manuscript and while it was being reviewed, a paper appeared in the literature \citep[hereafter A18]{anand2018} based on an analysis of the same \textit{HST} images presented here. The A18 authors assumed that the stars we identify as being RGB stars are really AGB stars---implying Coma P has an extreme SFH dominated by very recent star formation---and concluded that Coma P is located at a distance of 10.9 $\pm$ 1.0 Mpc.

We reanalyzed our CMD in an attempt to verify the suggested distance from A18.  We first used our CMD to search for a TRGB at or near the location highlighted in A18.  We did not find any feature in our CMD that matches with the putative TRGB identified by A18; our CMD presented in Figure \ref{fig:CMD_background} shows no noticeable concentration of stars at F814W $\sim$ 26.12, the location identified by A18 for their alternative TRGB.

Next, we downloaded the photometric data file used in the A18 paper from their website, and compared it with ours.  We found numerous examples of stars that were present in their CMD but not in ours.  In particular, there are many stars with F606W $-$ F814W $<$ 0.8 and F814W brighter than 26.0 (i.e., in the portion of the CMD where photometric errors are low) that appear in Figure 1 of A18 that are not present in our  Figure \ref{fig:CMD_background} (left panel).  Stars in this color and magnitude range are present in the CMD of our background fields.  We have identified the specific stars that are missing from our CMD in this portion of the diagram, and (1) verified that we did actually detect them, and (2) identified the reason they are not in our CMD.  In all cases, these stars were either outside of the area we identified as being in the galaxy using the method described in Section \ref{sec:obs} (see Figure \ref{fig:annuli_test}), or were rejected because they failed the photometric sharpness and/or crowding criteria we used.  We obtained the photometry filtering parameters applied in A18 from the first author (G. S. Anand 2018, private communication) who verified that they did not use the same sharpness values and did not apply any crowding cuts. If we remove all crowding cuts from our photometric data, we find a modest discontinuity in the F814W LF at 26.3 mag.   We believe that this is a photometric artifact caused by poorly recovered point sources in the data.

We also explored the proposal made by A18 that the stars above their TRGB magnitude of 26.12 are AGB stars that are 0.5$-$1 Gyr old. This population of stars reaches more than 2.5 mag brighter than their identified TRGB and appears at similar colors. To test the plausibility of this scenario, we simulated a series of CMDs using the PARSEC stellar evolution library.  We adopted the A18 distance of 10.9 Mpc and scaled our derived stellar mass for that distance (1.7 $\times$ 10$^6$ $M_\odot$).  Adopting the A18 suggestion for a significantly higher SFR between 0.5 and 1 Gyr ago produces a population of AGB stars that are only $\sim0.5$ magnitude brighter than the TRGB with F606W $-$ F814W colors of 1.5$-$2.5, significantly fainter and redder than the population of stars in the Coma~P CMD. 

We explored different SFH parameters in an attempt to create a CMD with any age population of AGB stars that matches the CMD of Coma~P at the A18 distance. Although not physically motivated, we found the closest approximation to the upper CMD of Coma~P assumes a burst of star formation occurring from 100 to 150 Myr ago that is 10$\times$ higher than the average SFR. However, while this SFH produces a similar number of stars in the upper CMD, the brightest AGB stars are still only 1.5 mag brighter than the TRGB. Furthermore, these bright AGB stars are also the reddest, reaching F606W $-$ F814W colors of 2.5, significantly offset from the color of the actual CMD. This is not surprising given that brighter AGB stars are known to be redder due to their pulsation properties and mass loss, even in low-metallicity dwarfs \citep{Jackson2007, Boyer2009, Boyer2015a, Boyer2015b}.  The nearly complete lack of observed stars redder than F606W $-$ F814W = 1.5 is in direct conflict with the suggestion that the stars brighter than F814W = 26 mag are AGB stars, and can be taken as further evidence against the A18 hypothesis.

To summarize, we were unable to verify the distance proposed by A18 unless we include point sources that are poorly recovered in the photometry. We also find that if we adopt the larger proposed distance, we are unable to reproduce a population of stars that A18 claim to be AGB stars in the magnitude and color range of the point sources in the CMD of Coma P. We conclude that our distance measurement of 5.5 Mpc is strongly preferred over the distance proposed  by A18.

\section{Discussion}\label{sec:discussion}

\subsection{Properties of Coma P}\label{sec:properties}

Observed and derived properties of Coma P are listed in Table~\ref{tab:Parameters}.  
The positional coordinates of Coma P have been updated slightly based on our \textit{HST} image, relative to the values given in \citet{steven2015}.  The integrated magnitudes are derived using the ground-based WIYN images of \citet{steven2015}, but using apertures that match those used in this paper (e.g., Figure~\ref{fig:final_region}) as well as employing more appropriate object masking based on the \textit{HST} images.  This resulted in significantly brighter magnitudes (e.g., 0.36 mag brighter in \textit{g}) compared to the original \citet{steven2015} photometry.    The values for the \ion{H}{1} data are taken directly from the ALFALFA survey \citep[e.g.,][]{haynes2011}.   These latter numbers are in good agreement with the new VLA and WSRT synthesis observations presented by \citet{Ball2018}.

\begin{deluxetable}{lccc}
\tabletypesize{\footnotesize}
\tablewidth{0pt}
\tablecaption{AGC 229385  -- Observed and Derived Properties.}
\tablehead{\colhead{Observed Quantities}&&\colhead{Value}&}
\startdata
R.A. (J2000)				&		&	12:32:10.9		&	\\
decl. (J2000)				&		&	+20:25:23.0		&	\\
$m_{g}$					&		&	18.84 $\pm$ 0.07	&	\\
$m_{r}$					&		&	18.86 $\pm$ 0.05	&	\\
$m_{i}$					&		&	19.00 $\pm$ 0.24	&	\\
(\textit{g}$-$\textit{r})$_o$				&		&	-0.05 $\pm$ 0.09       &      \\
$m_{B}$\tablenotemark{a}&		&	19.06 $\pm$ 0.08	&	\\
(\textit{B}$-$\textit{V})$_o$\tablenotemark{a}	&		&	 0.19	$\pm$ 0.11        &      \\
$A_{F606W}$             &       &    0.077        & \\
$A_{F814W}$             &       &    0.048        & \\
Angular diameter\tablenotemark{b} (arcsec)	&		&	78 $\pm$ 5	&   \\	
Ellipticity             &          &    0.54       & \\
Position angle ($\degr$) &          &    157 & \\
\noindent
$V_{\odot}$\tablenotemark{c} (km s$^{-1}$)		&		&	1348 $\pm$ 1	&	\\
$W_{50}$\tablenotemark{c} (km s$^{-1}$)		&		&	34 $\pm$ 1	&	\\
$S_{{HI}}$\tablenotemark{c} (Jy km s$^{-1}$)	&		&	4.87	$\pm$ 0.04	&	\\
\hline
\noindent Derived Quantities	&		&	Value	&	\\
\hline
(\textit{m} $-$ \textit{M})					&		&	28.70 $^{+0.11}_{-0.21}$		&		\\
Distance  (Mpc)			&		&	5.50 $^{+0.28}_{-0.53}$		&		\\
$M_{B}$					&		&	$-$9.75 $^{+0.13}_{-0.22}$		&		\\
Diameter (pc)			&		&	2080 $^{+170}_{-240}$			&		\\
\ion{H}{1} mass ($M_{\odot}$)	&		&	3.48\e{7}		&		\\
Stellar mass ($M_{\odot}$)	&	&		&		\\
SFH model:		&		&	4.7$\pm1.0$\e{5}		&		\\
M/L \textit{estimate:}		&		&	3.9\e{5}		&		\\
{\it Average:}		&		&	4.3\e{5}		&		\\
$M_{HI}$/$M_*$\tablenotemark{d}	&		&	81			&                \\
$M_{HI}$/$L_B$ ($M_{\odot}$/$L_{\odot}$)	&		&	28			&                
\enddata
\label{tab:Parameters}
\tablenotetext{a}{The \textit{B} magnitude and \textit{B $-$ V} color are derived from our measured \textit{g} \& \textit{r} magnitudes (see text) using the conversion relations presented in {\it http://www.sdss3.org/dr8/algorithms/sdssUBVRI
Transform.php\#Lupton2005}.}
\tablenotetext{b}{Major-axis diameter; see \S 2.}
\tablenotetext{c}{All \ion{H}{1} parameters taken from the ALFALFA database \citep[e.g.,][]{haynes2011}.}
\tablenotetext{d}{$M_{HI}$/$M_*$ is calculated using the average of the two stellar mass measurements.}
\end{deluxetable}

Based on our analysis and discussion in the previous section, we adopt the TRGB distance of 5.50 $^{+0.28}_{-0.53}$ Mpc for the distance to Coma~P.   Throughout the remainder of this paper this distance will be used for all analyses.  We note that this new distance is substantially lower than the value adopted by \citet{steven2015}.  In that study, a distance of 25 Mpc was assumed, based on the flow-model distance of \citet{masters2005}. The reduction in distance by a factor of $\sim$4.5 combined with our revised photometric measurements dramatically changes many of the derived physical parameters.  For example, the \textit{B}-band absolute magnitude is reduced from $-$12.72 to $-$9.75.  The major-axis diameter reduces to 2.08 $^{+0.17}_{-0.24}$ kpc, measured at an SB of $\sim$27.0 mag/arcsec$^2$ in the \textit{r}-band (i.e., a much lower level than is typically used for reporting isophotal diameters).  The new estimate for the \ion{H}{1} mass is 3.48 $\times$ 10$^7$ $M_\odot$.   If we take $M_{HI}$ to be a good approximation for $M_{baryon}$, then Coma P no longer sits nearly two orders of magnitude above the BTFR as reported in \citet{steven2015}, although it is still located substantially above the BTFR of \citet{mcnichols2016} \citep[e.g., see Figure 17 of][]{Ball2018}.  Finally, the change in distance results in a decrease in the inferred stellar mass.  Based on simple mass-to-light ratio estimators \citep[e.g.,][]{bell2001,bell2003,mcgaugh2014}, we use our total aperture photometry to derive an estimate of the stellar mass of 3.9 $\times$ 10$^5$ $M_\odot$.  Similarly, the SFH modeling presented in \S\ref{sec:sfh} yields an estimate of $M_*$ of 4.7$\pm1.0\times$ 10$^5$ $M_\odot$. We adopt $4.3 \times 10^5$, the average of the two measurements, as the stellar mass of Coma~P. 

While the revised distance has reduced the size, mass, and luminosity of Coma P, it remains an outlier in the distance-independent parameters such as $M_{HI}$/$M_*$ and $M_{HI}$/$L_B$.  Based on our new photometric measurements and the stellar mass estimates reported above, the \ion{H}{1} gas mass to stellar mass ratio is 81. Likewise, the $M_{HI}$/$L_B$ ratio is 28.0.  This value is much higher than those found in more normal galaxies, which typically have $M_{HI}$/$L_B$ between 0.15 and 4.2 \citep{roberts1994,stil2002,lee2003}.  Coma P is an exceptional system in terms of its \ion{H}{1} properties.

Few galaxies have properties similar to Coma P.  One galaxy that is reasonably similar in terms of its optical characteristics is the very low-mass, low-metallicity galaxy Leo P.  Coma P is slightly more luminous and larger than Leo P \citep[which has $M_B$ = $-$8.9 and diameter = 1.0 kpc;][]{rhode2013,mcquinn2015b}.  However, the two systems differ markedly in terms of their gaseous properties.   Coma P has an \ion{H}{1} mass more than 40 times larger than that of Leo P.  Leo P has an $M_{HI}$/$M_*$ ratio of only 1.4, a factor of 58 times smaller than Coma P.  Hence, even though Leo P was touted as being exceptionally gas-rich when it was discovered, its \ion{H}{1} characteristics are dwarfed by those of Coma P. 

Modest tension remains between the very blue ground-based photometric colors and the colors of the brightest stars plotted in our CMD.  As shown in Table~\ref{tab:Parameters}, the reddening-corrected (\textit{g$-$r}) and (\textit{B$-$V}) colors are quite blue, especially for a system that is not currently forming new stars.  As seen in \citet[][see their Figure 2]{steven2015}, Coma P is also quite prominent in GALEX far-UV imaging, also consistent with the observed blue colors.  On the other hand, the CMD shown in Figure~\ref{fig:CMD_background} is dominated by red giant stars, with only a handful of blue main-sequence stars present above F814W = 27.0.  The composite color of the resolved stars in the CMD corresponds to F606W$-$F814W = 0.80 (corresponding roughly to (\textit{r$-$i}) $\sim$ 1.0), far redder than the global photometry.  However, due to the extremely low SB nature of Coma~P, the global photometry is quite uncertain.   Our analysis reveals that small variations in the background measurements can lead to large (0.2 -- 0.3 mag) changes in the magnitudes and colors.  Hence, our ``formal" photometric errors are likely to be lower limits.  Despite the apparent tension between integrated photometric colors and composite color of the resolved stars, the recent SFRs calculated from the far-ultraviolet (FUV) emission and the CMDs are in agreement with each other to within the uncertainties (see \S\ref{sec:sfh}).

\begin{figure*}[ht]
\epsscale{0.51}
\plotone{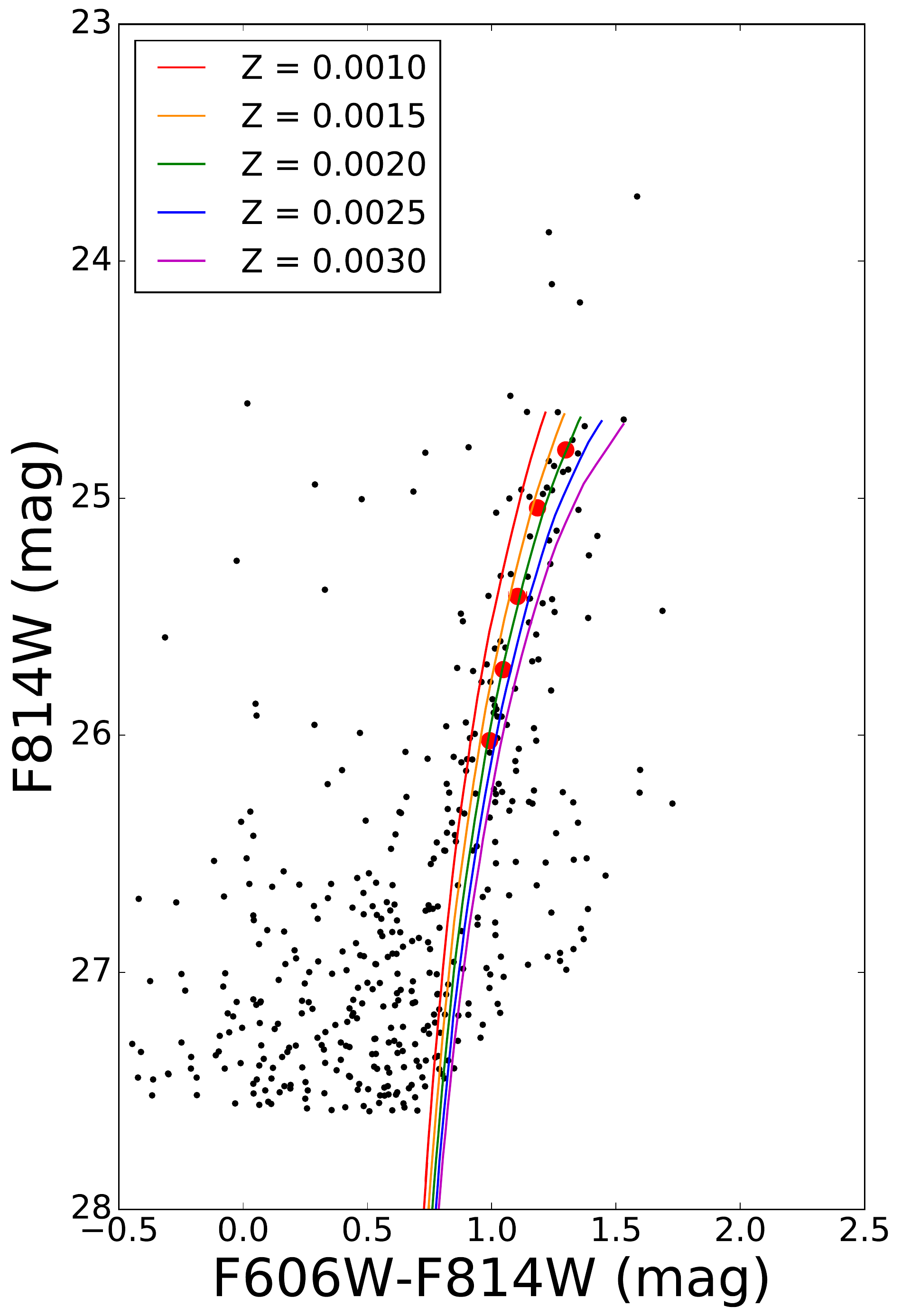}
\caption{CMD of Coma P overplotted with PARSEC isochrones for a 10 Gyr population with metallicities ranging from \textit{Z}~=~0.0010 to \textit{Z}~=~0.0030 in increments of 0.0005.  We binned the data along the RGB because it was difficult to estimate the metallicity with the scatter in the RGB.  The average locations of the RGB stars are plotted as large red circles. The binned RGB points appear to lie along the \textit{Z}~=~0.0020 isochrone with a scatter of 0.0005.}
\label{fig:CMD_isochrones}
\end{figure*}

\subsection{Model Isochrones and Metallicity Estimate}\label{sec:isochrones}

In order to estimate the metallicity of the stars in Coma P, isochrones with varying metallicities were overlaid on the CMD.  Padova PARSEC, MIST, and Dartmouth isochrones were compared to see if there was a sizable difference between stellar models. 

\subsubsection{PARSEC Models}\label{sec:parsec}

\begin{figure*}[ht]
\epsscale{0.59}
\plotone{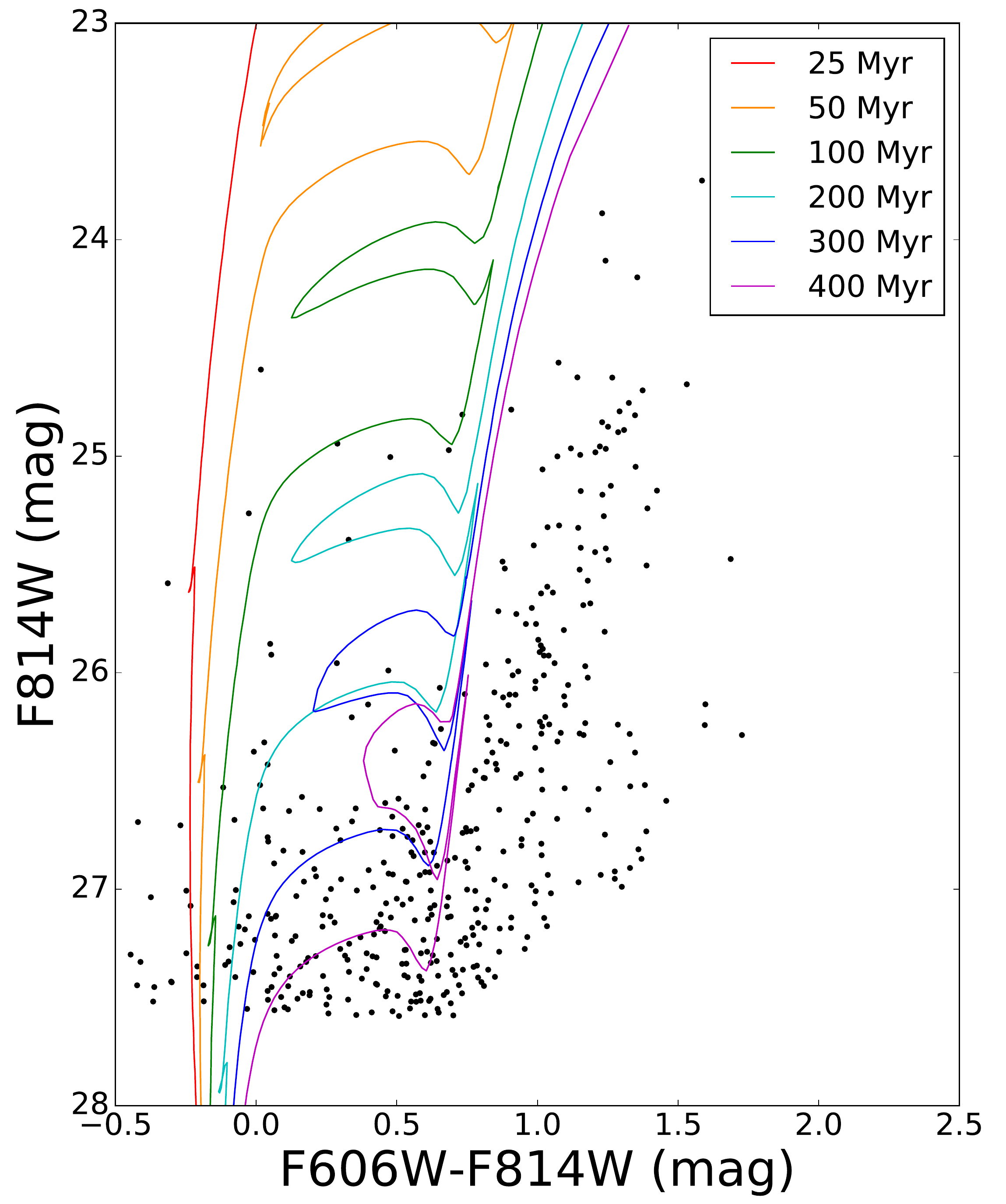}
\caption{CMD of Coma P overplotted with PARSEC isochrones (\textit{Z} = 0.0020) with ages ranging from 25 to 400 Myr.  There is a small group of stars that could be the main sequence and helium-burning stars of a 100-200 Myr stellar population (the green and cyan isochrones).  A few objects that could be associated with a younger ($\sim$50$-$100 Myr) population of main-sequence stars are also present, although there is no well-defined upper main sequence present in Coma P.}
\label{fig:young_isochrones}
\end{figure*}

The PARSEC models computed by the Padova group \citep{bressan2012,marigo2017} use up-to-date stellar models and offer many choices in metallicity to the user.  We generated PARSEC isochrones in the \textit{HST} ACS filters with no dust and a foreground Galactic absorption of $A_{V}$~=~0.086 \citep{Schlafly2011}.  After comparing isochrones of a single metallicity and ages ranging from 2 to 10 Gyr, we chose to use isochrones with an age of 10 Gyr.  The isochrones of different ages exhibited a small spread in the location of the RGB, but the differences were negligible for ages above 5 Gyr.  The lack of a distinct upper main sequence indicated that the bulk of the stellar population was older, which also informed our choice of 10 Gyr.

After choosing a single age for the stellar population, isochrones with metallicities of 0.0010~$\leqslant$~Z~$\leqslant$~0.0030 ($\Delta Z$=0.0005) were generated.  For comparison, the solar metallicity for the PARSEC isochrones is $Z_{\odot}$~=~0.0152.  These isochrones are plotted over the CMD of Coma P in Figure~\ref{fig:CMD_isochrones}.  The width of the RGB makes it more difficult to identify which isochrone fits the data the best.  Because of this scatter in the RGB, we binned the data along the RGB into five average points.  The RGB was first isolated in color and magnitude space, and then bins were chosen along the RGB so that each bin was the same size in $\Delta$F814W.  The average color of the stars in the bin and the average F814W magnitude were calculated for each bin, and the average points are plotted on the CMD in Figure~\ref{fig:CMD_isochrones} as large red circles along the RGB.  The red circles appear to lie along the \textit{Z}~=~0.0020 isochrone, which we adopt as our metallicity estimate for Coma~P.  This corresponds to log(O/H)~+~12~=~7.80 or [Fe/H]~= $-$0.88 (if [Fe/H]~=~[Z/H]).  The low-metallicity estimate ($\sim$1/8th solar\footnote{Assumes a solar oxygen abundance of 8.69 \citep{asplund2009}.}) is consistent with the dwarf nature of Coma~P.  This abundance is actually a bit higher than one would predict for a galaxy with the mass and luminosity of Coma~P using existing mass$-$metallicity (M-Z) or luminosity$-$metallicty (L-Z) relations \citep[e.g.,][]{Berg2012, Hirschauer2018}.  However, we stress that our method for estimating the metal content of the older stars in Coma~P makes a direct comparison with more accurately determined present-day nebular abundances difficult.

\subsubsection{Other Isochrone Models}\label{sec:othermods}

In addition to the PARSEC isochrones, we also analyzed the MIST and Dartmouth isochrones \citep[e.g.,][]{dotter2008,choi2016,dotter2016}.  These isochrone models differ from each other slightly in the metallicity input and also in the solar metallicity calibration.  Using the MIST model, we estimate the metallicity of the stellar population of Coma~P to be roughly \textit{Z} = 0.0013 or [Fe/H] = -1.05.  Using the Dartmouth model, we estimate the metallicity of the stellar system of Coma~P to be roughly [Fe/H] = -0.80. These bracket the metallicity estimate from the PARSEC isochrones.  We adopt the metallicity from the PARSEC isochrone models of \textit{Z}~=~0.0020 with an uncertainty of 0.0005 as the metallicity estimate of Coma~P.  

\subsubsection{Young Age Isochrones}\label{sec:youngage}

In addition to using isochrones for estimating the metallicity of the stars in the galaxy, we also overlaid young isochrones generated using the PARSEC code to identify possible young stars in the galaxy's CMD.  Figure~\ref{fig:young_isochrones} shows the CMD for Coma P with isochrones of various ages overlaid.  The ages used range from 25 to 400 Myr, and all of the isochrones have a metallicity of \textit{Z} = 0.0020.  The lower age limit was chosen to be consistent with the observed lack of any H$\alpha$ emission in Coma P.  There is a small group of stars that are consistent with a helium-burning sequence of a stellar population with an age of approximately $100-200$ Myr. Although we cannot rule out the possibility that some of these point sources are background or foreground contaminants, it is unlikely that all of them are contaminants given the lack of sources at F814W $\sim$25 mag in a background region of equal area shown in Figure~\ref{fig:CMD_background}. This analysis is in agreement with the SFH computed for Coma P  in the following Section~\ref{sec:sfh}.\\

\subsection{SFH of Coma P}\label{sec:sfh}

\begin{figure*}[ht]
\epsscale{0.5}
\plotone{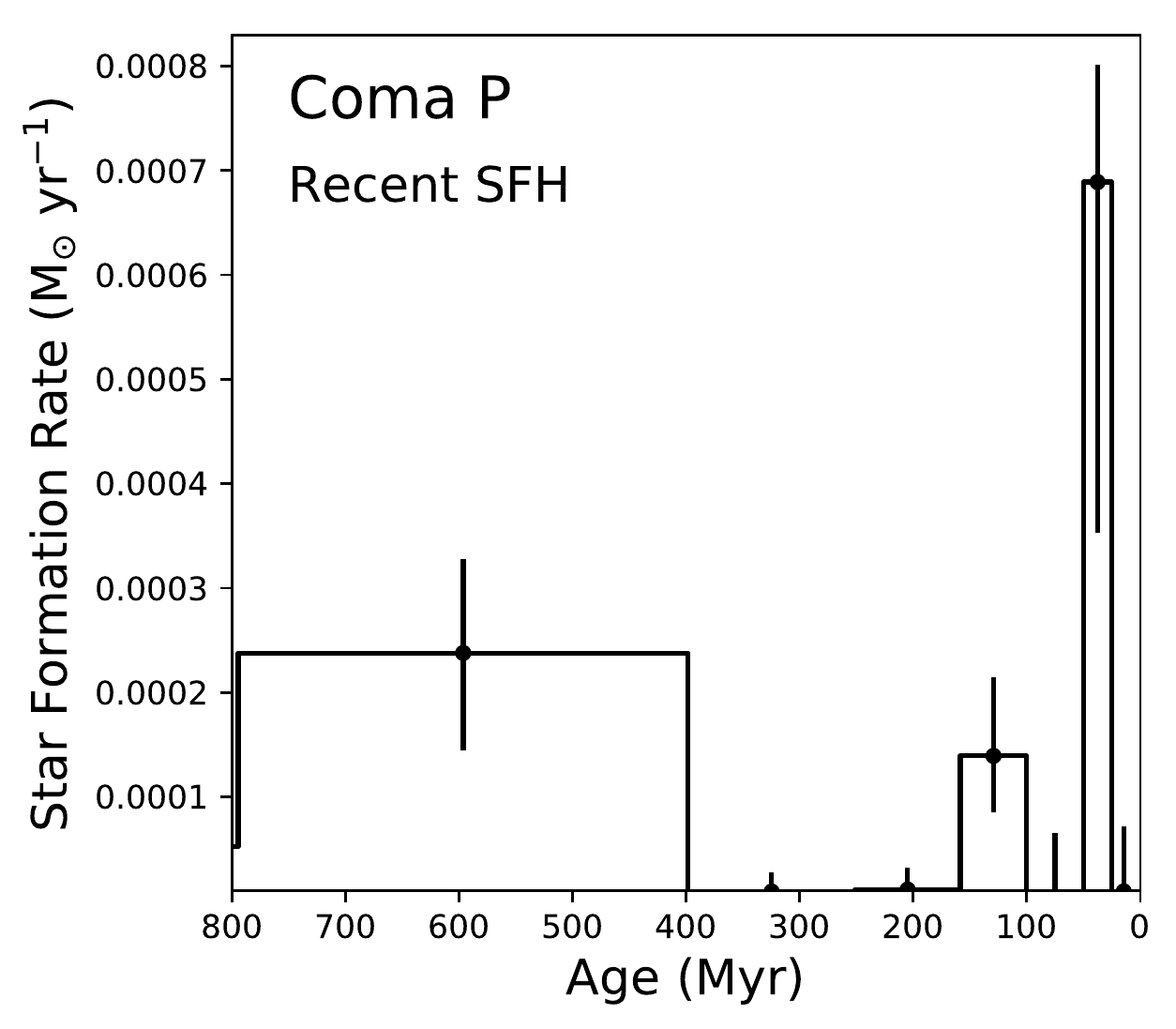}
\caption{Recent star formation history of Coma P. Star formation has been intermittent with short episodes of activity. Comparing the recent SFH with the total stellar mass of Coma~P, 15\% of the stars were formed in the last 800 Myr.}
\label{fig:sfh}
\end{figure*}

The SFH of Coma P was reconstructed from the resolved stellar populations using the CMD-fitting routine \textsc{match} \citep{dolphin2002}. The technique uses the photometry, artificial stars, a stellar evolution library, and an assumed initial mass function (IMF) to create a series of synthetic simple stellar populations with varying ages and metallicities. This series of simple stellar populations is combined until the modeled CMD represents the best fit to the observed CMD based on a Poisson likelihood statistic. The corresponding SFH represents the most likely SFH of the galaxy. We refer the reader to \citet{dolphin2002} for more details on \textsc{match} and to \citet{tolstoy2009} for a review of CMD-fitting techniques.

To derive the SFH of Coma P, we used the updated Padova stellar evolution library PARSEC \citep{bressan2012} and assumed a Kroupa IMF \citep{kroupa2001}. The distance was fixed to the TRGB distance of 5.50 Mpc. We assumed a binary fraction of 35\% and Galactic foreground absorption from \citet{Schlafly2011}, and we restricted the metallicity to increase as a function of time. 

From the best-fitting SFH, the total stellar mass formed over the lifetime of the galaxy is $8.2^{+1.2}_{-1.6}\times10^{5}$ $M_\odot$. Assuming a return mass fraction of 42\% for a Kroupa IMF \citep{Vincenzo2016}, the current stellar mass in Coma P is $4.7\pm1.0\times10^5$ $M_\odot$. This is in good agreement with the stellar mass estimate of $3.9\times10^5$ $M_\odot$ based on the mass-to-light ratio (see \S\ref{sec:properties}). Given the gas-rich nature of Coma~P, the overall low stellar mass suggests an extremely inefficient star formation process over the history of the galaxy. The best-fitting metallicity corresponds to an isochrone metallicity of \textit{Z} = 0.0038, in general agreement with our fit from the overlying isochrones in \S\ref{sec:isochrones}. 

The temporal resolution achievable in a derived SFH depends on both the photometric depth of the data and the number of stars in a CMD. For a well-populated CMD, the recent SFH is well constrained with time bins as short as 25 Myr from data of modest photometric depth (i.e., $\sim2$ mag below the TRGB) as there is sufficient information on the ages of young stars in the upper main sequence and helium-burning sequences \citep{mcquinn2012}. However, accurately resolving the ancient SFH of a galaxy with high temporal resolution ($\delta t\sim1$ Gyr) requires a well-populated CMD and photometry reaching below the old main-sequence turnoff, which has been achieved only in Local Group galaxies with HST observations. For a full discussion of the effects of photometric depths and measuring SFHs at different lookback times, see \citet{mcquinn2010}. 

Given the depth of the CMD, we focus primarily on the recent SFH of Coma~P. Figure~\ref{fig:sfh} shows the best-fitting SFH over the past 800 Myr.  Based on extensive tests with \textsc{match}, we determined the time-binning scheme of $\sim$25, 50, 100, 150, 250, 400, and 800 Myr, which resulted in robustly measured recent SFRs. Uncertainties include both statistical uncertainties determined through a hybrid Markov Chain Monte Carlo approach \citep{dolphin2013} and systemic uncertainties estimated using Monte Carlo simulations that shift the stellar evolution models in both bolometric luminosity and temperature and re-solve for the SFH \citep{dolphin2012}.

As seen in Figure~\ref{fig:sfh}, Coma~P has experienced intermittent star formation over the last 800 Myr. There has been little activity in the past 25 Myr preceded by a more significant star forming episode from 25 to 50 Myr ago. Integrating the stellar mass formed most recently and comparing that with the stellar mass formed over the lifetime of the galaxy, Coma~P has formed 15\% of its stars over the past 800 Myr compared with 85\% over the first $\sim 12$ Gyr. The recent SFRs are higher than a constant SFR model. It is unclear from the current data whether the stellar mass buildup prior to 800 Myr occurred primarily at early times followed by a period of extended quiescence, or if star formation has occurred at an approximately constant rate. 

Hydrodynamical simulations using cold dark matter favor early star formation scenarios in very low-mass galaxies, which can be temporarily quenched by stellar feedback and reionization. In such models, reignition of star formation can be triggered by the cooling of a galaxy's extended gas reservoir or by an environmental influence \citep[e.g.,][]{fitts2018, wright2018}. By contrast, hydrodynamical simulations using warm dark matter favor delayed onset of star formation in the mass range of Coma~P, with little to no star formation occurring at the earliest epochs \citep{Bozek2018}, similar to the delayed mass assembly seen in the low-mass galaxy KDG~215 \citep{Cannon2018}. Observations of Coma~P reaching below the old main-sequence turn-off are needed to explore these SFH scenarios; such observations will be possible with the upcoming \textit{James Webb Space Telescope}.

\begin{figure*}[ht]
\epsscale{0.50}
\plotone{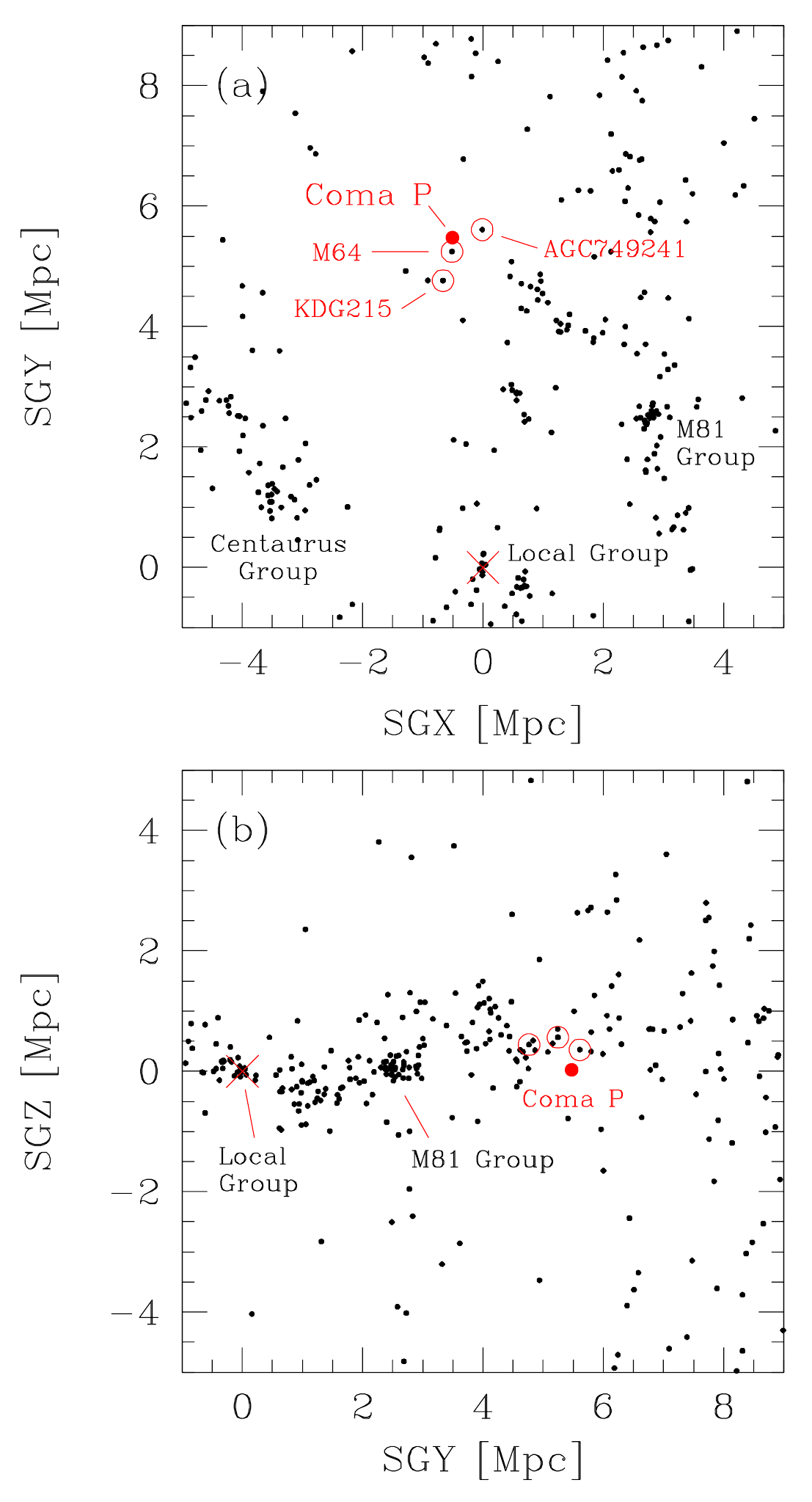}
\caption{Large-scale structure around Coma P plotted in supergalactic coordinates. (Top) The SXG$-$SGY plane, looking down onto the supergalactic plane. (Bottom) The SGY$-$SGZ plane, presenting a side view of the supergalactic plane. Coma P is located very close to the plane.  Only three galaxies lie within 1 Mpc of Coma P: NGC 4826 (M64), AGC 749241, and KDG 215.}
\label{fig:environment}
\end{figure*}

The SFR averaged over the last 100 Myr from the resolved stellar populations can be compared with the SFR measured from the integrated FUV light. \citet{steven2015} measured the FUV flux from GALEX images of Coma P and reported an SFR based on the integrated light\footnote{The NUV and FUV SFRs presented in \citet{steven2015} were inadvertently reversed in Table 2 of that paper.  The value listed for the FUV SFR corresponds to the NUV value, and vice versa.}, an assumed velocity-based distance of 25 Mpc, and the scaling relations from \citet{murphy2011} and \citet{hao2011}. Here, we update this SFR using the TRGB distance of 5.50 Mpc and the first empirically derived FUV SFR scaling relation using CMD-fitting techniques from \citet{mcquinn2015a}, converted to a Kroupa IMF by scaling down by a factor of 0.67. We find the SFR$_{FUV} = (3.1\pm1.8) \times 10^{-4}$ $M_{\odot}$ yr$^{-1}$, slightly higher than the CMD-based SFR averaged over the most recent 100 Myr of $(1.7\pm0.8) \times 10^{-4}$ $M_{\odot}$ yr$^{-1}$, but still within the uncertainties of the measurements. The FUV scaling relation is predicated on the SFRs being constant over the past 100 Myr, where the birth and death of FUV bright stars are approximately equal. Given the variability in the SFR seen in Fig.~\ref{fig:sfh}, it is not surprising that the integrated light measurements are slightly higher than the CMD-based method based on counting the individual stars. 

\subsection{Environment around Coma P}\label{sec:environment}

\begin{figure*}[ht]
\epsscale{0.65}
\plotone{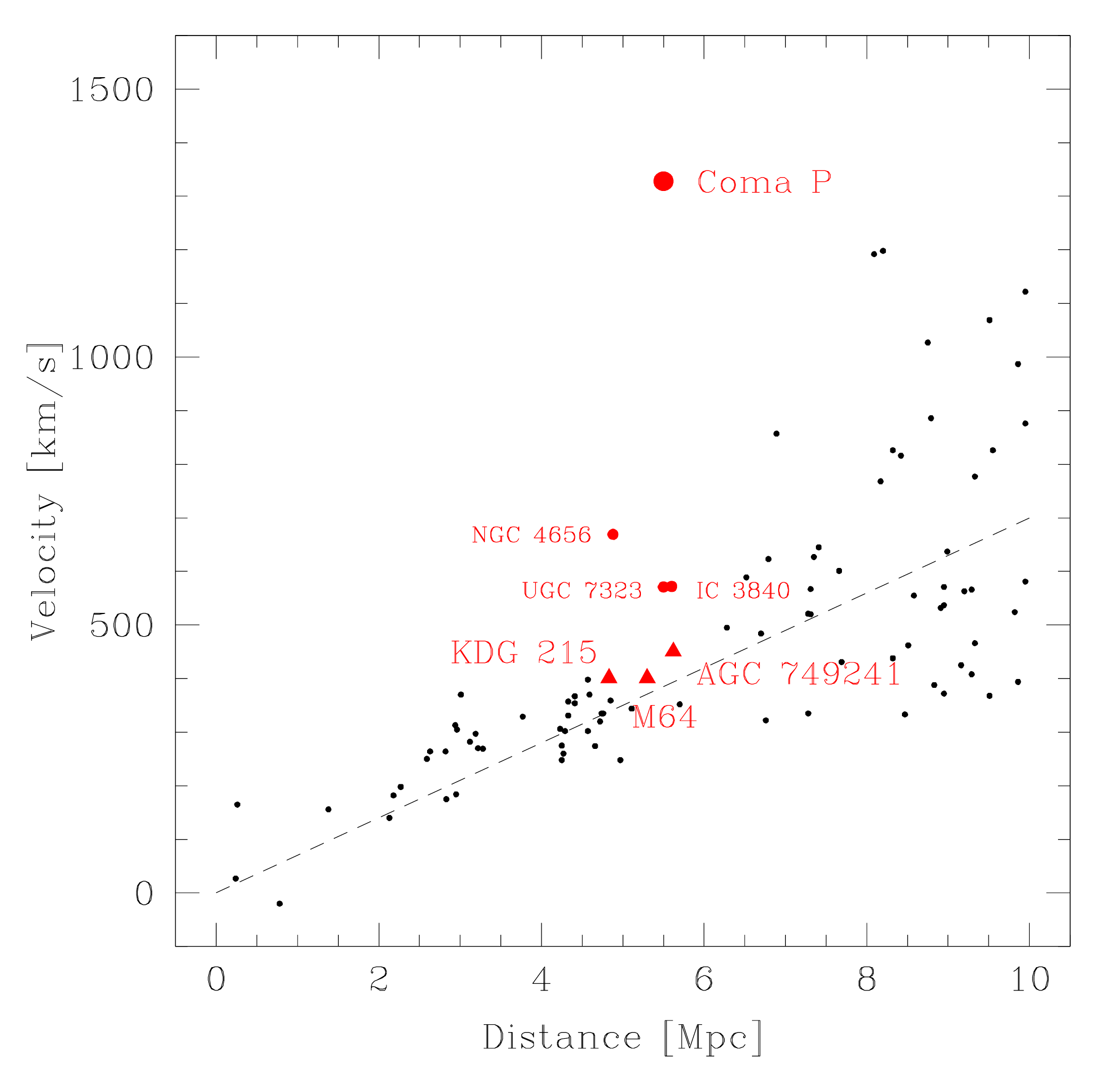}
\caption{Hubble diagram showing velocity versus distance for the same data set shown in Figure~\ref{fig:environment}. The dashed line represents a pure Hubble flow with $H_o$ = 70 km s$^{-1}$/Mpc. Coma P is shown as a red circle, and the three labeled galaxies closest to Coma P in Figure~\ref{fig:environment} are shown as red triangles.  Three galaxies with approximately the same distance as Coma P and positive peculiar velocities are marked with red points, though their peculiar velocities are not as extreme as Coma P.} 
\label{fig:hubblediag}
\end{figure*}

The \citet{steven2015} paper looked at the environment of the HI1232+20 system.  However, at that time it was assumed that Coma~P and its \ion{H}{1} companions were located at a distance of 25 Mpc.  Our newly derived distance to Coma~P of only 5.5 Mpc necessitates a new look at the local environment of this system.   To carry out this analysis, we utilize the most recent version of the Cosmicflows-3 \footnote{The data used for the environmental analysis presented in this section were obtained from the Extragalactic Distance Database: edd.ifa.hawaii.edu} database \citep{tully2016}.  This database tabulates velocity-independent distances derived via a range of methods.  At the distances relevant for comparison with Coma~P, the majority of the galaxies in the Cosmicflows-3 database have distances derived via the TRGB method.  Only a handful have distances that rely on the less accurate Tully$-$Fisher method.

Figure~\ref{fig:environment} plots the locations of nearby galaxies using Cartesian supergalactic coordinates (SGX, SGY, SGZ).  The Milky Way is located at the origin of this coordinate system (indicated by the large red X).  Coma~P is shown as a red dot in both panels.  The upper panel plots the SGX$-$SGY plane (i.e., looking down on the supergalactic plane).  The Local Group, as well as the nearby M81 and Centaurus groups, is labeled.  The lower panel of Figure~\ref{fig:environment} plots the SGY$-$SGZ plane.  In this plot the supergalactic plane becomes clearly evident.  Coma~P is located very close to this plane; it has coordinates SGX = -0.51 Mpc, SGY = 5.48 Mpc, and SGZ = 0.03 Mpc.

Based on the available data, Coma~P appears to be in a relatively low-density pocket despite being located right in the supergalactic plane.  The nearest galaxy to Coma~P with a distance in the Cosmicflows-3 database, NGC 4826 (M64), is located 590 kpc away.  This bright Sab spiral galaxy lies at distance of \textit{D} = 5.30 Mpc and has an absolute magnitude of $M_B$ = $-$19.5, slightly less luminous (and presumably slightly less massive) than the Milky Way.  Only slightly farther away is AGC~749241, a gas-rich dwarf galaxy that is part of the SHIELD sample \citep{Cannon2011,mcquinn2014}.  \citet{mcquinn2014} gave a TRGB distance of 5.62 Mpc, which places it 610 kpc away from Coma~P.   AGC~749241 has an absolute magnitude of $M_B$ = $-$10.25, comparable to that of Coma~P.  The only other galaxy within 1 Mpc of Coma~P is KDG~215.  This low SB dwarf galaxy ($M_B$ = $-$12.0) is 840 kpc from Coma~P\footnote{A recent reanalysis of the available HST-based CMD by \citet{Cannon2018}  arrives at a slightly larger distance for KDG 215 of 5.11 Mpc, compared to 4.83 Mpc in the Cosmicflows-3 database.  With this revised distance, the separation between Coma P and KDG 215 reduces to 650 kpc.  Our interpretation of the Coma P system does not change based on this modest revision.}.  These three galaxies are labeled in Figure~\ref{fig:environment}.  Based on the distances alone, one is inclined to conclude that Coma~P and the entire HI1232+20 system are evolving in isolation. 

The distances by themselves do not tell the full story, however.  The observed heliocentric velocity of Coma~P is 1348 km/s \citep{steven2015}, compared to 410 km/s for M64.  Hence, despite being separated by 590 kpc from M64, Coma~P has a line-of-sight (LOS) velocity of 938 km/s relative to this massive spiral.  The same situation is true for AGC~749241 and KDG~215: Coma P has a relative velocity of $\sim$900 km/s with respect to these two galaxies.  Figure~\ref{fig:hubblediag} shows a Hubble diagram of velocity vs. distance, using the same data that were used to create Figure~\ref{fig:environment}.   Only galaxies within 30\degr~of Coma~P on the sky are included in this figure.  The dashed line corresponds to a pure Hubble flow with $H_o$ = 70 km s$^{-1}$ Mpc$^{-1}$.  Coma~P, M64, AGC~749241, and KDG~215 are all labeled.  The large peculiar velocity of Coma~P relative to the other galaxies at this distance is immediately evident.

Assuming that the measured distances are correct (i.e., Coma~P and M64 are at essentially the same distance), the large velocity difference between the two is difficult to reconcile. It is highly unusual for galaxy pairs located in low-density environments to exhibit such large relative velocities.  Given the 5.86\degr~separation on the sky between Coma~P and M64, the minimum separation between the two is 560 kpc.  We note that since it is extremely unlikely that the entire relative motion of the two galaxies would be along the LOS, the true 3D relative velocity could be substantially higher than the 938 km/s radial velocity difference.   We also note that there are three other galaxies at roughly the same distance as Coma~P/M64 that also have significant positive peculiar velocities.   These galaxies are UGC 7323 and IC 3840, both at a distance of $\sim$5.50 Mpc and $V_{\rm Helio}$ $\sim$ 572 km/s ($V_{pec}$ $\sim$ 187 km/s), and NGC 4656, at a distance of 4.88 Mpc and with $V_{\rm Helio}$ = 669 km/s ($V_{pec}$ = 327 km/s).   Although the peculiar velocities of these galaxies are not as extreme as that of Coma~P, taken together, they may be indicating the presence of a minor coherent mass flow at this distance, to which Coma~P also belongs.  The previous study of \citet{kara2011} also hints at the existence of this region of higher peculiar velocities (see their Figure 4).

We discuss the large peculiar velocity of Coma P further in the following section.

\subsection{Proposed Scenario to Explain the Observed Properties of Coma~P}\label{sec:scenario}

The combined optical and \ion{H}{1} properties \citep[e.g.,][]{Ball2018} of Coma P are extreme.  In this section, we attempt to develop a plausible evolutionary scenario to explain the observed characteristics of this enigmatic galaxy.  The properties that our scenario must explain include (i) the extremely high \ion{H}{1} to stellar mass ratio ($M_{HI}$/$M_*$ = 81), (ii) the very large peculiar velocity relative to its closest neighbors ($V_{pec}$ $\sim$ 930 km/s), (iii) an SFH that includes an extremely inefficient star formation process given the available gas mass and a modest increase in star formation over the past $\sim$1 Gyr, and (iv) the unusual nature of the HI1232+20 system, which contains four distinct \ion{H}{1} clouds, only one of which contains a detected stellar population.

Perhaps the most important clue to disentangling the evolutionary history of Coma~P is the large peculiar velocity.  It seems unreasonable to assume that Coma~P has spent its entire existence in close proximity to its current nearest neighbors.   A peculiar velocity of 938 km/s relative to M64 corresponds to a relative motion of very nearly 1 Mpc per Gyr. 
Therefore, it would seem likely that Coma~P formed quite far from the vicinity of M64.

A plausible scenario is that this system was originally formed deep within the void located above the supergalactic plane (positive SGZ in Figure ~\ref{fig:environment}b).  As the local cosmic structures formed, it would have found itself initially quite isolated.  This would naturally lead to a fairly quiescent existence.   It most likely formed the majority of its stars during an initial formation process, then, due to a lack of perturbing interactions, existed for a substantial length of time with little or no subsequent star formation.  During this period, it could have been a classic low SB galaxy.  Due to its location within the local void, its substantial gas reservoir was hung up in a stable rotating configuration and very little of it was converted into stars.  This would result in a low SB galaxy with a very large gas-to-star mass ratio.  

Eventually, the gravitational pull of the surrounding higher density regions would have accelerated Coma~P away from the void center and toward the nearest void wall.  We hypothesize that Coma~P arrived in the vicinity of the supergalactic plane between 1 and 2 Gyr ago, crossing the plane in the vicinity of M64.  We imagine that the fast-moving gas-rich dwarf system had a high-speed fly-by encounter with M64 approximately 1 Gyr ago.  The large mass difference between the two galaxies would have resulted in a large acceleration of Coma P relative to M64.  Most of the currently observed peculiar velocity of Coma P would result from this large-mass-ratio ($\sim$5000:1) encounter, rather than from its motion due to the influence of the large-scale matter distribution.  
As a result of this encounter, much of the outer gas disk of Coma~P was shredded, while the centrally located stellar population remained largely intact.  Some of the gas from Coma~P might have been accreted by M64.  It has been known for some time that M64 possesses a gas disk with two counter-rotating components \citep{braun1992,braun1994}.  It has generally been assumed that the counter-rotating disk in M64 is due to the merger of a gas-rich galaxy with the spiral galaxy.  However, our scenario involving a high-speed interaction with a system like Coma~P would also be consistent with the observed nature of M64.

After the encounter, Coma~P would have continued past the large spiral, moving in a direction that would take it toward negative values of SGZ (i.e., below the supergalactic plane) and larger positive values of SGY (away from the Milky Way, to account for its positive peculiar velocity).  The encounter may well have resulted in some gas compression in the inner portions of the galaxy, igniting star formation.  At this point, the outer gas disk of Coma~P would have been severely disrupted.  However, as time passes, much of that gas would tend to clump and potentially fall back onto the parent galaxy.  This appears to be the phase at which we are seeing Coma~P today.  The  starless components of the HI1232+20 system represent gas that was flung outward from the central body as part of a high-speed interaction.   Their \ion{H}{1} masses of 9.7 $\times$ 10$^6$ $M_\odot$ (for AGC~229384) and 5.8 $\times$ 10$^6$ $M_\odot$ (for AGC~229383) represent $\sim$30\% of the cold neutral gas in the HI1232+20 system.  Whether these larger clumps will ever fall back onto Coma~P is unclear.  The detailed \ion{H}{1} map presented by \citet{Ball2018} suggests that another large clump of \ion{H}{1} gas is currently falling back into Coma~P, as evidenced by the irregular gas motions in the northern portion of the galaxy.  The episodic return of gas to the parent galaxy over time might well result in modest star-formation events occurring at irregular intervals in the time since the M64 encounter, as suggested by our best-fit SFR model shown in Figure 8.  This re-initiation of star formation has been noted in hydrodynamical simulations of low-mass galaxies after encounters with gas thrown off by galaxy interactions \citep{wright2018}.

The combined optical and \ion{H}{1} properties of Coma~P are extreme.  The evolutionary scenario we have presented appears to explain most of the key observational characteristics of Coma~P as well as of the HI1232+20 system.  It even addresses the older puzzle of the counter-rotating gas disks in M64.  We do not claim that this scenario is the correct explanation for Coma~P; undoubtedly, additional scenarios could be developed that also cover the observational facts.  Our goal is to hypothesize a plausible explanation for how Coma~P and the HI1232+20 system evolved.  We suggest that the next step would be to utilize cosmological simulations to test whether the type of interaction described in our scenario can result in a low-mass dwarf being accelerated to the level necessary to acquire the large peculiar velocity observed.
If our scenario is correct, then it implies that Coma~P provides a rare opportunity to study the physical processes involved in \lq\lq void clearing\rq\rq\ at the extreme low-mass end.  Presumably, this process is fairly common, but the difficulty in discovering and observing very low-mass galaxies like Coma~P has, until now, limited our exposure to such extreme cases.

\subsection{Possible Connection to the Nearby (Almost) Dark System AGC 229360/61}\label{sec:neighbor}

Finally, we mention the possible connection between Coma~P and another pair of (almost) dark objects detected by ALFALFA.  AGC~229360 and AGC~229361 are located 1.4 degrees away from Coma~P on the sky and have velocities of 1573 km/s and 1661 km/s, respectively.   Like the \ion{H}{1} sources in the HI1232+20 system, these two \ion{H}{1} clouds both appear to lack a stellar component in SDSS images.  Recent ground-based WIYN images have revealed a very low SB stellar component associated with AGC~229361 (H. Pagel 2018, private communication).   It is tempting to associate this pair of \ion{H}{1} clouds with Coma~P and the HI1232+20 system due to their proximity on the sky as well as their similar velocities.    For example, it is possible that the AGC~229360/61 pair could be following a similar trajectory to Coma~P out of the local void.   At this time, however, any such connection would be built on circumstantial evidence; the true distance to AGC~229360/61 is not known.  If this galaxy pair were at a distance similar to Coma P, they would be exhibiting an even larger peculiar velocity of $\sim$1250 km/s relative to the other nearby galaxies.   Although AGC~229361 has a velocity 313 km/s higher than that of Coma~P, it is within 223 km/s of the galaxy AGC~222741, a background galaxy located in the HI1232+20 field \citep{steven2015}.   Hence, it is probably equally likely that the AGC~229360/61 pair is located farther away and is associated with AGC~222741.   Further observations of these systems would clearly be fruitful.

\section{Conclusions}\label{sec:conclusion}

We present optical imaging of Coma~P obtained with the \textit{HST} and use the data to derive a TRGB distance to this object.  Our resolved stellar photometry allows us to generate a CMD for Coma P, which shows a clear RGB as well as a small population of upper main-sequence stars.   We measure the distance to Coma~P to be 5.50$^{+0.28}_{-0.53}$ Mpc.  This is much closer than the original distance of 25 Mpc assumed from the velocity flow model used by ALFALFA \citep{steven2015}.   We consider possible alternative distances to Coma~P, but conclude that the bulk of the evidence supports our TRGB distance.
By fitting PARSEC isochrones to the CMD, we estimate the metallicity of Coma P to be \textit{Z}=0.0020 assuming an age of $\sim$10 Gyr for the RGB stars. 

The smaller derived distance to Coma~P changes the inferred properties of the galaxy significantly.   Our updated masses and luminosities are all dramatically lower than previously reported:  Coma~P is an extreme dwarf system.   Although it is no longer such a severe outlier on galaxy scaling relations \citep[e.g., the BTFR;][]{Ball2018},  it is still an enigmatic system.  For example, it has an \ion{H}{1} mass to blue luminosity ratio $M_{HI}$/$L_{B}$ of 28, much higher than the values found in more normal galaxies, and a hydrogen gas-to-star ratio $M_{HI}$/$M_*$ of 81. We have also investigated the SFH of Coma~P.  Using a CMD-fitting technique, we found that the best-fit SFH shows episodic star formation over the past 800 Myr, representing 15\% of the mass formed over the lifetime of the galaxy.

The new distance estimate to Coma P compelled us to investigate the local environment of the system.  Despite being in the supergalactic plane, Coma P is in a relatively low-density environment, with the closest galaxy being approximately 590 kpc away.  Coma P has a large peculiar velocity relative to the other galaxies in the vicinity, which may help explain its complicated evolutionary scenario.  We put forth the hypothesis that Coma P formed in the void located above the supergalactic plane as a low SB galaxy with a large gas-to-star mass ratio.  Over time, the gravitational pull of the galaxies located on the edge of the void accelerated Coma P toward the nearest void wall.   Upon entering the supergalactic plane, Coma P interacted with M64 in a high-speed fly-by that severely disrupted its outer gas disk and triggered new star formation within Coma~P.  Some of the gas from Coma~P was accreted onto M64, creating its observed counter-rotating gas disk, while large clumps of \ion{H}{1} gas continued to travel with Coma~P past M64 to their present location, where we observe the HI1232+20 system to consist of Coma~P and three starless \ion{H}{1} clouds that together contain $\sim$30\% of the gas mass of the system.

\acknowledgements
Support for program GO-14108 was provided by NASA through a grant from the Space Telescope Science Institute, which is operated by the Association of Universities for Research in Astronomy, Inc., under NASA contract NAS 5-26555.  S.W.B. and J.J.S. gratefully acknowledge financial support from the College of Arts and Sciences at Indiana University.  We wish to thank the support staff of the Space Telescope Science Institute for their expert help during all phases of this project.  J.M.C. and C.B. acknowledge support from NSF grant AST-1211683.  S.J. acknowledges support from the Australian Research Council’s Discovery Project funding scheme (DP150101734).  K.L.R. is supported by NSF Astronomy \& Astrophysics grant AST-1615483.  The ALFALFA team at Cornell is supported by NSF grants AST-0607007, AST-1107390, and AST-1714828 and by grants from the Brinson Foundation. 

\facility{\textit{Hubble Space Telescope}}

\end{document}